\documentclass{jfm}
\usepackage{amsmath,stmaryrd,natbib,upgreek}

\begin{document}

\newtheorem{lemma}{Lemma}
\newtheorem{corollary}{Corollary}

\shorttitle{Electrohydrodynamics of viscous drops in strong electric fields} 
\shortauthor{D. Das and D. Saintillan} 

\title{Electrohydrodynamics of viscous drops in strong electric fields: Numerical simulations}

\author
 {
 Debasish Das\footnote{Present address:  Department of Applied Mathematics and Theoretical Physics, Centre for Mathematical Sciences, University of Cambridge, Wilberforce Road, Cambridge CB3 0WA, UK}
\and
 David Saintillan
 \corresp{\email{dstn@ucsd.edu}},
 }
   
\affiliation
{
Department of Mechanical and Aerospace Engineering, University of California San Diego, \\ 9500 Gilman Drive, La Jolla, CA 92093, USA
}

\maketitle

\begin{abstract}
Weakly conducting dielectric liquid drops suspended in another dielectric liquid and subject to an applied uniform electric field exhibit a wide range of dynamical behaviors  contingent on field strength and material properties. These phenomena are best described by the  Melcher--Taylor leaky dielectric model, which hypothesizes charge accumulation on the drop-fluid interface and prescribes a balance between charge relaxation, the jump in Ohmic currents from the bulk and charge convection by the interfacial fluid flow. Most previous numerical simulations based on this model have either neglected interfacial charge convection or restricted themselves to axisymmetric drops. In this work, we develop a three-dimensional boundary element method for the complete leaky dielectric model to systematically study the deformation and dynamics of liquid drops in electric fields. The inclusion of charge convection in our simulations permits us to investigate drops in the Quincke regime, in which experiments have demonstrated a symmetry-breaking bifurcation leading to steady electrorotation. Our simulation results show excellent agreement with existing experimental data and small-deformation theories.
\end{abstract}

\vspace{-1.5cm}

\section{Introduction}

The dynamics and deformations of immiscible liquid droplets suspended in another fluid medium and subject to an electric field find a wide range of applications in industrial processes, including ink-jet printing \citep{basaran2013}, electrospinning \citep{huang2003}, oil extraction from oil-water emulsions \citep{schramm1992,
eow2002}, electrospraying and atomization of liquids \citep{taylor1964,taylor1969,castellanos2014} and microfluidic devices and pumps \citep{stone2004,laser2004}. Their study is also important in understanding natural phenomena such as electrification of rain, bursting of rain drops in thunderstorms and electrification of the atmosphere \citep{simpson1909, blanchard1963}. Of interest to us in this work is the case of dielectric liquids such as oils, which are poor conductors. Unlike aqueous electrolytes, where the dynamics arises from the action of the electric field on diffuse Debye layers extending into the liquid bulk, these so-called leaky dielectric liquids are typically characterized by the absence of bulk charges; any net charge in the system instead concentrates at interfaces between liquid phases as a result of the mismatch in material properties. Dynamics and deformations then result from the action of the field on this surface charge, which induces interfacial stresses and can drive fluid flows.  

\begin{table}
  \begin{center}\vspace{-0.4cm}
  \begin{tabular}{ccc}
  	 	 \textit{Experimental work}: & & \\[2pt]
   		 \citet{allan1962,torza1971,vizika1992,tsukada93}; & & \\[2pt]
   		 \citet{krause1998,ha2000a,ha2000b,sato2006}; & & \\[2pt]
   		 \citet{salipante2010,salipante2013,karyappa2014,lanauze2015}. & & \\[5pt]
    		 \textit{Theoretical modeling} (EHS):  & & \\[2pt]
   		 \citet{konski1953,harris1957}; \\[2pt]
   		 \citet{allan1962,taylor1964}. \\[5pt]
     	 \textit{Numerical simulation} (EHS): & & \\[2pt] 									 \citet{brazier1971a,brazier1971b,miksis1981}; & & \\[2pt]
     	 \citet{haywood1991,dubash2007a,dubash2007b}. & &\\[5pt]
    		 \textit{Theoretical modeling} (EHD): & & \\[2pt]
    		 \citet{taylor1966,torza1971,ajayi1978,esmaeeli2011}; \\[2pt]
	     \citet{zhang2013,lanauze2013,he2013,yariv2016}; & & \\[2pt]
	     \citet{bandopadhyay2016,yarivalmog16,das2016}. & & \\[5pt]
     	 \textit{Numerical simulation} (EHD): & & \\[2pt]	                   				 \citet{sherwood1988,feng1996,baygents1998,feng1999}; \\[2pt]
     	 \citet{hirata2000,lac2007,supeene2008,bjorklund2009}; \\[2pt]
     	 \citet{lopez2011,karyappa2014,hu2015,lanauze2015}. \\[5pt]
	    	 \textit{Reviews}: & & \\[2pt] 
	    	 \citet{melcher1969,saville1997,vlahovska2016}. \\[2pt]
    	   \end{tabular}
  \caption{Non-exhaustive summary of the literature on the deformations and dynamics of uncharged liquid drops subject to a uniform DC electric field. We distinguish electrohydrostatic models (EHS), which neglect fluid flow, from electrohydrodynamic models (EHD), where fluid flow is taken into account. } \label{summary}
  \end{center}
\end{table}

 We focus in this work on the simple case of an isolated leaky dielectric drop suspended in a weakly conducting liquid subject to a uniform DC electric field. This prototypical problem has fascinated scientists for decades and a summary of the existing literature on this problem is presented in table~\ref{summary}. Early studies in the field primarily focused on the specific cases of an either insulating or perfectly conducting drop suspended in an insulating fluid medium. In these cases, the drop-fluid interface does not experience any tangential electric stresses, and as a consequence fluid motions are absent and the drop can only attain a steady prolate shape as a result of a jump in electric pressure across the interface \citep{konski1953,harris1957}. Oblately deformed drops were first observed in experiments by \citet{allan1962}, suggesting an inconsistency in the existing electrohydrostatic models. In his pioneering work, \citet{taylor1966} realized that dielectric liquids, while poor conductors, still have a weak conductivity and can therefore carry free charges to the drop-fluid interface. The action of the electric field on these surface charges then gives rise to tangential electric stresses that generate toroidal circulatory currents now known as Taylor vortices. By incorporating this effect into a small-deformation theory, Taylor was able to predict both prolate and oblate shapes depending on material properties, and his results compared favorably with experiments.

The discovery of these surface charges and their role in generating fluid motions motivated \citet{melcher1969} to develop a more complete framework for studying the electrohydrodynamics of leaky dielectric drops. The cornerstone of their work is a surface charge conservation equation that prescribes a balance between transient charge relaxation, the jump in Ohmic currents from both bulk fluids and charge convection along the drop surface due to the interfacial fluid flow. Taylor's original theory based on this model accounted for first-order deformations in the limit of vanishing electric capillary number $Ca_E$, denoting the ratio of electric to capillary forces. While predicted deformation values showed good agreement with experimental results \citep{torza1971} in weak fields where deformations are small, significant departures were observed with increasing field strength. In an attempt to resolve this discrepancy, \citet{ajayi1978} calculated drop deformations to second order in $Ca_E$, yet his results did not improve upon Taylor's solution in the case of oblate drops when compared with experiments. This systematic mismatch was a consequence of the neglect of nonlinear interfacial charge convection in these  models. There have since then been numerous attempts to extend these original predictions by including additional  effects such as transient shape deformation \citep{haywood1991,esmaeeli2011}, transient charge relaxation \citep{zhang2013}, fluid acceleration \citep{lanauze2013}, interfacial charge convection \citep{feng2002,shkadov02,he2013,das2016}, and sedimentation \citep{bandopadhyay2016,yarivalmog16}. 

Various numerical schemes have also been developed over the years to address this problem computationally. \citet{brazier1971a}, \citet{brazier1971b} and \citet{miksis1981} used the boundary element method to solve the electrohydrostatics problem, wherein the shape of the drop is evolved quasi-statically so as to balance normal stresses on the interface. In a more comprehensive study, \citet{sherwood1988} solved the coupled electrohydrodynamic problem assuming creeping flow conditions, which allowed him to use the boundary element method for both the electric and flow problems. His pioneering work was extended by \citet{baygents1998} to study axisymmetric drop pair interactions and by \citet{lac2007} to investigate a much wider range of electric and fluid parameters. Very recently, \citet{lanauze2015} extended these models by formulating an axisymmetric boundary element method for the complete Melcher--Taylor leaky dielectric model. Other methods based on finite elements \citep{feng1996,feng1999,hirata2000,supeene2008}, level sets \citep{bjorklund2009}, the immersed boundary method \citep{hu2015} and the volume-of-fluid method \citep{lopez2011} have also been employed to investigate drop dynamics. 

Recent experiments, however, have uncovered another dynamical regime in strong electric fields \citep{krause1998,ha2000b, sato2006,salipante2010}. Upon increasing field strength, a symmetry-breaking bifurcation has been reported in the case of weakly conducting drops, by which the axisymmetric shape predicted by the aforementioned models becomes unstable and gives rise to a non-axisymmetric tilted drop configuration accompanied by a rotational flow. In yet stronger fields, chaotic dynamics have also been reported, with unsteady stretching and tumbling of the drop \citep{salipante2013}, sometimes leading to breakup \citep{ha2000b}. This curious transition, most recently described in the work of \citet{salipante2010,salipante2013}, shares similarities with the electrorotation of weakly conducting rigid particles in strong electric fields, which is well known since the work of \citet{quincke1896} and has been explained in detail theoretically \citep{jones1984,das2013}. The case of a deformable drop, however, is significantly more challenging than that of a rigid particle, due to the deformations of the interface and to the complexity of the interfacial flow, which does not follow rigid body dynamics. Theoretical models for Quincke electrorotation of droplets are scarce and have all assumed a spherical shape as well as weak \citep{he2013} or strong \citep{yariv2016} charge convection by the flow. Computational models are non-existent to our knowledge, as nearly all simulation methods developed in the past have only allowed for axisymmetric shapes, which is sufficient to describe the oblate and prolate deformations arising in weak fields but is inadequate to capture symmetry breaking. A notable exception is the work of  \citet{lopez2011}, who simulated the electrohydrodynamics of three-dimensional drops using the volume-of-fluid approach but did not address the Quincke regime. 

In this work, we develop three-dimensional boundary element simulations of the electrohydrodynamics of a liquid droplet based on a formulation for the complete Melcher--Taylor leaky dielectric model. This enables us to investigate dynamics both in the axisymmetric Taylor regime of weak fields as well as in the Quincke regime of strong fields; to our knowledge, these are the first numerical simulations to capture Quincke electrorotation of drops in three dimensions. Our numerical results show excellent agreement with both existing experimental data and small-deformation theories. Details of the boundary integral formulations for the electric and flow problems and their numerical implementations are described in \S \ref{sec:BIF} as well as in the appendices. Simulation results and comparisons with previous experiments and theories are discussed in \S \ref{sec:results}. We conclude by summarizing our work and discussing possible extensions in \S \ref{sec:conclusion}.

\section{Problem definition}\label{sec:probdef}

\subsection{Governing equations}\label{sec:govern}

\begin{figure}
\begin{center}
\includegraphics[width=0.7\linewidth]{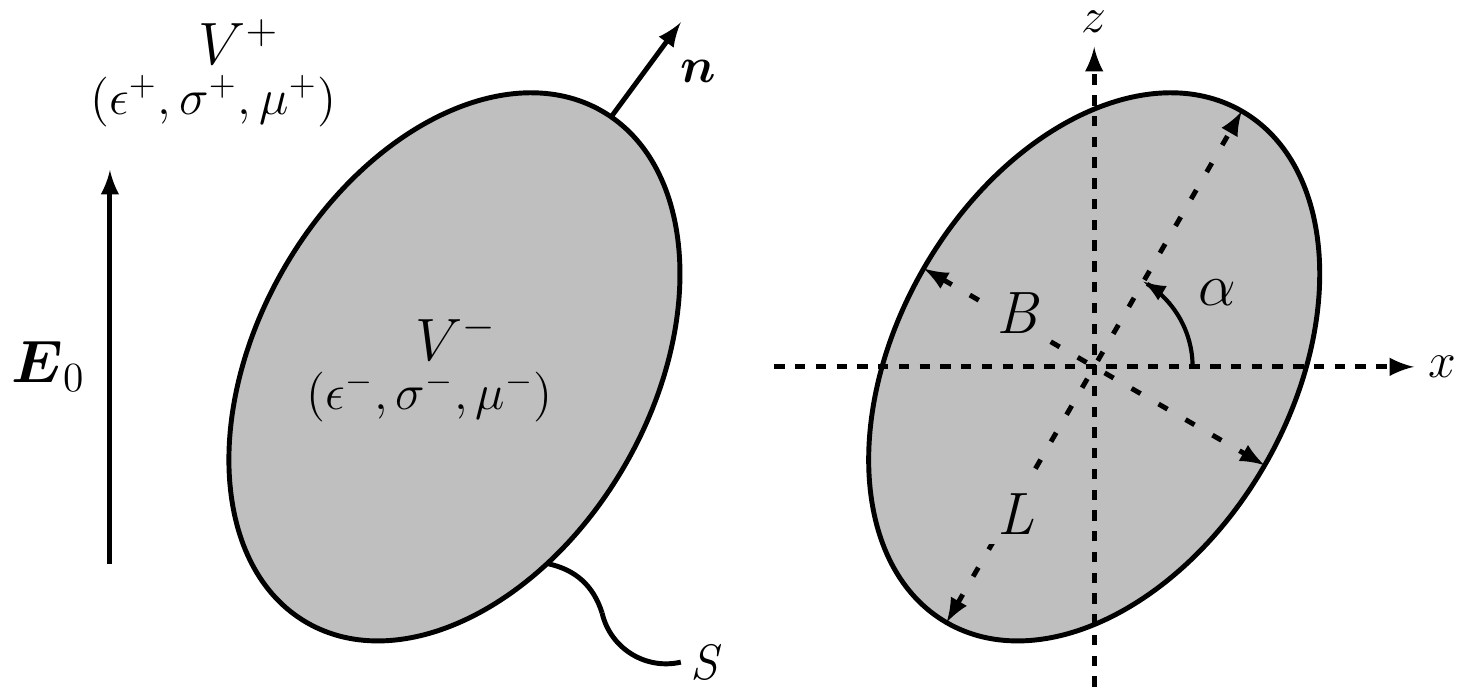}
\end{center}
\caption{Problem definition: A liquid droplet with surface $S$ and outward unit normal $\boldsymbol{n}$ is suspended in an unbounded domain and placed in a uniform electric field $\boldsymbol{E}_{0}$ pointing in the vertical direction. $V^{\pm}$ denote the exterior and interior domains, respectively, and $(\epsilon^\pm, \sigma^\pm,\mu^\pm)$ are the corresponding dielectric permittivities, electric conductivities and dynamic viscosities. The drop's major and minor axis lengths are denoted by $L$ and $B$, and the major axis is tilted at an angle $\alpha$ with respect to the horizontal direction.}
\label{fig:figure1}
\end{figure}

We consider an uncharged neutrally buoyant liquid droplet with undeformed radius $a$ occupying volume $V^{-}$ in an infinite fluid medium $V^{+}$ and subject to a uniform electric field $\boldsymbol{E}_{0}$ as depicted in figure~\ref{fig:figure1}. The drop surface is denoted as $S$ and has an outward unit normal $\boldsymbol{n}$. Let $(\epsilon^{\pm},\sigma^{\pm},\mu^{\pm})$ be the dielectric permittivities, electric conductivities, and dynamic viscosities of the exterior and interior fluids, respectively. In the Melcher--Taylor leaky dielectric model \citep{melcher1969}, all charges in the system are concentrated on the drop surface, so that the electric potential in both fluid domains is harmonic:
\begin{equation}
\nabla^{2}\varphi^{\pm}(\boldsymbol{x})=0 \qquad \mbox{for}\,\,\,\boldsymbol{x}\in V^{\pm}. \label{eq:laplace}
\end{equation}
On the drop surface, the electric potential is continuous, as is the tangential component of the local electric field:
\begin{align}
\llbracket \varphi (\boldsymbol{x})\rrbracket =0 \quad \mbox{and} \quad \llbracket \boldsymbol{E}_t(\boldsymbol{x})\rrbracket =\boldsymbol{0} \qquad \mbox{for}\,\,\, \boldsymbol{x}\in S,
\end{align}
where $\boldsymbol{E}^{\pm}_{t}=({\mathsfbi{I}}-\boldsymbol{nn})\bcdot \boldsymbol{E}^{\pm}$ and $\boldsymbol{E}^{\pm}=-\bnabla \varphi^{\pm}$. We have introduced the notation $\llbracket f(\boldsymbol{x})\rrbracket \equiv f^+(\boldsymbol{x}) - f^-(\boldsymbol{x})$ for any field variable $f(\boldsymbol{x})$ defined on both sides of the interface. Unlike $\boldsymbol{E}_t$, the normal component of the electric field $E_{n}^{\pm}=\boldsymbol{n}\bcdot\boldsymbol{E}^{\pm}$ undergoes a jump due to the mismatch in electrical properties between the two media \citep{landau1984}, which results in a surface charge distribution $q(\boldsymbol{x})$ related to the normal displacement field by Gauss's law:
\begin{equation}
q(\boldsymbol{x})=\llbracket \epsilon {E}_n(\boldsymbol{x})\rrbracket \qquad \mbox{for}\,\,\, \boldsymbol{x}\in S.
\end{equation}
The surface charge density $q$ evolves due to two distinct mechanisms: Ohmic currents from the bulk and advection by the fluid flow with velocity $\boldsymbol{v}(\boldsymbol{x})$ on the drop surface. Accordingly, it satisfies the conservation equation:
\begin{equation}
\partial_{t}q + \llbracket \sigma {E}_n\rrbracket +\bnabla_{s}\bcdot (q\boldsymbol{v})=0 \qquad \mbox{for}\,\,\, \boldsymbol{x}\in S, \label{eq:chargeeq0}
\end{equation}
where $\bnabla_{s}\equiv ({\mathsfbi{I}}-\boldsymbol{nn})\bcdot\bnabla$ is the surface gradient operator. On neglecting unsteady terms and surface charge convection, equation \eqref{eq:chargeeq0} reduces to the simpler boundary condition $\llbracket \sigma {E}_n\rrbracket=0$ used in a number of previous studies \citep{sherwood1988,baygents1998,lac2007}.

The fluid velocity field $\boldsymbol{v}^{\pm}(\boldsymbol{x})$ and corresponding pressure field $p^{H\pm}(\boldsymbol{x})$ satisfy the Stokes equations in both fluid domains:
\begin{equation}
-\mu^{\pm}\nabla^{2}\boldsymbol{v}^{\pm}+\bnabla p^{H\pm}=\boldsymbol{0}\quad \mbox{and}\quad \bnabla\bcdot\boldsymbol{v}^{\pm}=0 \qquad \mbox{for}\,\,\,\boldsymbol{x}\in V^{\pm}. 
\end{equation}
The velocity is continuous on the drop surface:
\begin{equation}
\llbracket \boldsymbol{v}(\boldsymbol{x})\rrbracket =\boldsymbol{0} \qquad \mbox{for}\,\,\, \boldsymbol{x}\in S,
\label{eq:kinematic}
\end{equation}
and, in the absence of Marangoni effects, the jumps in electric and hydrodynamic tractions across the interface balance interfacial tension forces:
\begin{equation}
\llbracket \boldsymbol{f}^{E}\rrbracket + \llbracket \boldsymbol{f}^{H}\rrbracket =\gamma (\bnabla_{s}\bcdot\boldsymbol{n})\boldsymbol{n} \qquad \mbox{for}\,\,\, \boldsymbol{x}\in S.\label{eq:stressbalance}
\end{equation}  
Here, $\gamma$ is the constant surface tension and $\bnabla_{s} \bcdot\boldsymbol{n}=2\kappa_{m}$ is twice the mean surface curvature.  The jumps in tractions are expressed in terms of the Maxwell stress tensor ${\mathsfbi{T}}^{E}$ and hydrodynamic stress tensor ${\mathsfbi{T}}^{H}$ as
\begin{align}
\llbracket \boldsymbol{f}^{E}\rrbracket &=\boldsymbol{n}\bcdot\llbracket{\mathsfbi{T}}^{E}\rrbracket=\boldsymbol{n}\bcdot\llbracket \epsilon (\boldsymbol{EE}-\tfrac{1}{2}E^{2}{\mathsfbi{I}})\rrbracket, \\
\llbracket \boldsymbol{f}^{H}\rrbracket &=\boldsymbol{n}\bcdot\llbracket{\mathsfbi{T}}^{H}\rrbracket = \boldsymbol{n}\bcdot \llbracket-p^H{\mathsfbi{I}}+\mu\left(\bnabla \boldsymbol{v}+\bnabla\boldsymbol{v}^{T}\right)\rrbracket.
\end{align}
The jump in electric tractions can also be expressed as
\begin{align}
\begin{split}
\llbracket \boldsymbol{f}^{E}\rrbracket =\llbracket \epsilon E_{n} \rrbracket  \boldsymbol{E}_{t}+\tfrac{1}{2}\llbracket \epsilon(E_{n}^{2}-E_{t}^{2})\rrbracket \boldsymbol{n} 
=q\boldsymbol{E}_{t}+ \llbracket p^{E}\rrbracket \boldsymbol{n}.\label{eq:electraction}
\end{split}
\end{align}
 The first term on the right hand side captures the tangential electric force on the interface arising from the action of the tangential field on the interfacial charge distribution. The second term captures normal electric stresses and can be interpreted as the jump in an electric pressure $p^E=\frac{1}{2}\epsilon (E_n^2-E_t^2)$ \citep{lac2007}.
 
\subsection{Non-dimensionalization}\label{sec:nondim}
 
Non-dimensionalization of the governing equations yields five dimensionless groups, three of which are ratios of material properties typically defined as:
 \begin{equation}
 R=\frac{\sigma^+}{\sigma^-}, \qquad  Q= \frac{\epsilon^-}{\epsilon^+},\qquad \lambda =\frac{\mu^-}{\mu^+}.  
 \end{equation}
The low-drop-viscosity limit $\lambda\rightarrow 0$ describes a bubble, whereas $\lambda\rightarrow \infty$ describes a rigid particle. The product $RQ$ can also be interpreted as the ratio of the inner to outer charge relaxation times: 
\begin{equation}
RQ=\frac{\tau^-}{\tau^+}\qquad \mbox{where}\qquad \tau^+=\frac{\epsilon^+}{\sigma^+}, \quad \tau^-=\frac{\epsilon^-}{\sigma^-}.
\end{equation}
A possible choice for the two remaining dimensionless numbers consists of the electric capillary number $Ca_E$ and electric Mason number $Ma$ defined as
\begin{equation}
Ca_{E}=\frac{a\epsilon^+ E_{0}^{2}}{\gamma}, \qquad Ma = \frac{\mu^+}{\epsilon^+ \tau_{MW}E_{0}^2}. \label{eq:MaCa}
\end{equation}
The electric capillary number $Ca_E$ compares the characteristic time $\tau_{\gamma}$ for a deformed drop to relax to its equilibrium shape as a result of surface tension to the electro-viscous timescale $\tau_{EHD}$ \citep{salipante2010}, each defined as
\begin{equation}
\tau_{\gamma}=\frac{\mu^+(1+\lambda)a}{\gamma}, \qquad \tau_{EHD}=\frac{\mu^+(1+\lambda)}{\epsilon^+ E_{0}^{2}}.
\end{equation}
On the other hand, the Mason number $Ma$ is the ratio of $\tau_{EHD}$, multiplied by a factor of $(1+\lambda)^{-1}$, to the Maxwell-Wagner relaxation time
\begin{equation}
\tau_{MW}=\frac{\epsilon^- +2\epsilon^+}{\sigma^- +2\sigma^+},
\end{equation}
which is the characteristic timescale for polarization of the drop surface upon application of the field \citep{das2013}. $Ma$ is also directly related to the ratio of the electric field magnitude $E_0$ to the critical electric field $E_{c}$ for onset of Quincke rotation of a rigid sphere as
\begin{equation}
Ma=\frac{\overline{\epsilon}-\overline{\sigma}}{2}\left(\frac{E_{c}}{E_{0}}\right)^{2},
\end{equation}
where
\begin{equation}
\overline{\epsilon}=\frac{\epsilon^- -\epsilon^+}{\epsilon^- + 2\epsilon^+}, \quad \overline{\sigma}=\frac{\sigma^- -\sigma^+}{\sigma^- + 2\sigma^+}, \quad E_{c}=\sqrt{\frac{2\mu^+}{\epsilon^+ \tau_{MW}(\overline{\epsilon}-\overline{\sigma})}}. \label{eq:Ec}
\end{equation}
For a rigid sphere, Quincke rotation occurs when $E_0>E_c$, or $Ma< (\overline{\epsilon}-\overline{\sigma})/2$, thus necessitating the application of a strong electric field. For the critical electric $E_c$ to take on a real value, the condition $\overline{\epsilon}>\overline{\sigma}$, which is equivalent to $RQ>1$ or $\tau^+>\tau^-$, needs to be satisfied; this generally implies that the drop is less conducting than the suspending fluid. It is useful to note the direct correspondence between $Ma$ and the electric Reynolds number $Re_E$ defined by other authors \citep{lanauze2015,schnitzer2015}:
\begin{equation}
Ma =\frac{\tau^+/\tau_{MW}}{Re_{E} } \qquad \mbox{where} \qquad Re_E= \frac{\epsilon^{+}E_{0}^{2}}{\sigma^+\mu^+}.
\end{equation}
Finally, an additional dimensionless group can also be constructed by taking the ratio of the capillary time $\tau_\gamma$ and Maxwell-Wagner relaxation time $\tau_{MW}$ and is independent of field strength \citep{salipante2010}:
\begin{equation}
Ca_{MW}=\frac{\tau_{\gamma}}{\tau_{MW}}=\frac{\mu^+ (1+\lambda)a}{\gamma \tau_{MW}}=(1+\lambda)Ca_{E}Ma. \label{eq:CaMW}
\end{equation}
For a fixed set of material properties, varying $Ca_{MW}$ is equivalent to varying drop size $a$. In the remainder of the paper, we exclusively use dimensionless variables by scaling lengths with $a$, electric fields with $E_{0}$, and times with $\tau_{MW}$. In addition to $R$, $Q$ and $\lambda$, we primarily use $Ca_E$ and $Ma$ as dimensionless groups, though some of the results in \S \ref{sec:results} will also be shown in terms of $E_{0}/E_{c}$ and $Ca_{MW}$.

\section{Boundary integral formulation}\label{sec:BIF}

\subsection{Electric problem}\label{sec:electric}

The solution of Laplace's equation \eqref{eq:laplace} is best formulated using boundary integral equations \citep{jaswon1963,symm1963,pozrikidis2002}. Following previous studies in the field \citep{sherwood1988,baygents1998,lac2007, lanauze2015} we represent the potential in terms of the single-layer density $\llbracket {E}_{n}(\boldsymbol{x})\rrbracket$ as
\begin{equation}
\varphi(\boldsymbol{x}_{0})=-\boldsymbol{x}_0\bcdot \boldsymbol{E}_{0}+\oint_{S} \llbracket {E}_{n}(\boldsymbol{x})\rrbracket \, \mathcal{G}(\boldsymbol{x}_0;\boldsymbol{x})\,\mathrm{d}S(\boldsymbol{x}) \qquad \mbox{for}\,\,\,\boldsymbol{x}_{0}\in V^{\pm}, S. \label{eq:intpotential}
\end{equation}
Here, $\boldsymbol{x}_{0}$ is the evaluation point for the potential and can be anywhere in space, whereas $\boldsymbol{x}$ denotes the integration point which is located on the drop surface. The Green's function or fundamental solution of Laplace's equation in an unbounded domain is given by
\begin{equation}
\mathcal{G}(\boldsymbol{x}_{0};\boldsymbol{x})=\frac{1}{4\uppi r}\quad \mbox{where} \quad \boldsymbol{r}=\boldsymbol{x}_{0}-\boldsymbol{x}, \,\,\, r=|\boldsymbol{r}|.
\end{equation}
Note that equation (\ref{eq:intpotential}) is valid in both fluid phases as well as on the interface since the Green's function is continuous across $S$. The equation is weakly singular, however, when $\boldsymbol{x}=\boldsymbol{x}_{0}$, though the singularity can be removed analytically by introducing plane polar coordinates in the parametric plane defining the local surface \citep{pozrikidis2002}. Knowledge of the single-layer potential density  $ \llbracket {E}_{n}(\boldsymbol{x})\rrbracket$ on the interface therefore allows one to determine the electric potential anywhere in space by simple integration, which prompts us to seek an equation for $ \llbracket {E}_{n}(\boldsymbol{x})\rrbracket$ in terms of the surface charge density $q$. We first take the gradient of equation \eqref{eq:intpotential} to obtain an integral equation for the electric field in the fluid:
\begin{equation}
\boldsymbol{E}^{\pm}(\boldsymbol{x}_0)=\boldsymbol{E}_{0}-\oint_{S} \llbracket {E}_{n}(\boldsymbol{x})\rrbracket \bnabla_{0}\mathcal{G}(\boldsymbol{x}_0;\boldsymbol{x})\,\mathrm{d}S(\boldsymbol{x}) \quad \mbox{for}\,\,\,\boldsymbol{x}_{0}\in V^{\pm}.
\end{equation}
The derivative of the Green's function undergoes a discontinuity at the interface, which needs to be accounted for when the evaluation point is on the boundary \citep{pozrikidis2011}:
\begin{equation}
\boldsymbol{E}^\pm(\boldsymbol{x}_0)=\boldsymbol{E}_{0}-\oint_{S} \llbracket {E}_{n}(\boldsymbol{x})\rrbracket \bnabla_{0}\mathcal{G}(\boldsymbol{x}_0;\boldsymbol{x})\,\mathrm{d}S(\boldsymbol{x})\pm \tfrac{1}{2}\llbracket {E}_{n}(\boldsymbol{x}_0)\rrbracket \boldsymbol{n}(\boldsymbol{x}_0) \quad \mbox{for}\,\,\,\boldsymbol{x}_{0}\in S.
\label{eq:intelectric}
\end{equation}
The integral equation for the electric field is strongly singular. However, taking a dot product on both sides with the unit normal $\boldsymbol{n}(\boldsymbol{x}_0)$ reduces the singularity by one order. Averaging the normal components of the  field outside and inside the drop then yields
\begin{equation}
\tfrac{1}{2}[E_n^+(\boldsymbol{x}_0)+E_n^-(\boldsymbol{x}_0)]=E_{n0}-\oint_{S} \llbracket {E}_{n}(\boldsymbol{x})\rrbracket \{\boldsymbol{n}(\boldsymbol{x}_0)\boldsymbol{\cdot}\bnabla_{0}\mathcal{G}(\boldsymbol{x}_0;\boldsymbol{x})\}\,\mathrm{d}S(\boldsymbol{x}) \quad \mbox{for}\,\,\,\boldsymbol{x}_{0}\in S,
\label{eq:intnormalsing}
\end{equation}
where the weak singularity can now be removed analytically following \citet{sellier2006}  by subtracting $\llbracket {E}_{n}(\boldsymbol{x}_0)\rrbracket$ from the single-layer density:
\begin{align}
\begin{split}
&\tfrac{1}{2}[E_n^+(\boldsymbol{x}_0)+E_n^-(\boldsymbol{x}_0)]+\llbracket {E}_{n}(\boldsymbol{x}_0)\rrbracket \left[\tfrac{1}{2}-L(\boldsymbol{x}_0)\right] \\
&=E_{n0}-\oint_{S} \{\llbracket {E}_{n}(\boldsymbol{x})\rrbracket-\llbracket {E}_{n}(\boldsymbol{x}_0)\rrbracket\}\{\boldsymbol{n}(\boldsymbol{x}_0)\boldsymbol{\cdot}\bnabla_{0}\mathcal{G}(\boldsymbol{x}_0;\boldsymbol{x})\}\,\mathrm{d}S(\boldsymbol{x}) \quad \mbox{for}\,\,\,\boldsymbol{x}_{0}\in S.
\label{eq:intnormalreg}
\end{split}
\end{align}
The scalar function ${L}(\boldsymbol{x}_0)$ is a purely geometrical quantity depending on drop shape and expressed as \citep{sellier2006}
\begin{align}
L(\boldsymbol{x}_0) = \boldsymbol{n}(\boldsymbol{x}_0)\bcdot  \oint_S \Big\{ [\boldsymbol{\nabla} \mathcal{G} \bcdot \boldsymbol{n}(\boldsymbol{x})] [\boldsymbol{n}(\boldsymbol{x})-\boldsymbol{n}(\boldsymbol{x}_0)] + \mathcal{G}(\boldsymbol{x}_0;\boldsymbol{x}) [\boldsymbol{\nabla}\bcdot \boldsymbol{n}] (\boldsymbol{x}) \boldsymbol{n}(\boldsymbol{x})\Big\} \,\mathrm{d}S(\boldsymbol{x}). \label{eq:integralL}
\end{align}
Gauss's law also allows us to express $E_n^+$ and $E_n^-$ on each side of the interface in terms of the jump in normal electric field,
\refstepcounter{equation}
$$
E_n^+=\frac{q-Q\llbracket {E}_{n}\rrbracket}{1-Q}, \qquad \qquad {E}_n^-=\frac{q-\llbracket {E}_{n}\rrbracket}{1-Q}, \eqno{(\theequation{\mathit{a},\mathit{b}})} \label{eq:normalelectric}
$$
which, after substitution into equation (\ref{eq:intnormalreg}), provides a regular integral equation for the jump $\llbracket {E}_{n}\rrbracket$: \vspace{-0.2cm}
\begin{align}
\begin{split}
&\oint_{S} \{\llbracket {E}_{n}(\boldsymbol{x})\rrbracket-\llbracket {E}_{n}(\boldsymbol{x}_0)\rrbracket\}\{\boldsymbol{n}(\boldsymbol{x}_0)\boldsymbol{\cdot}\bnabla_{0}\mathcal{G}(\boldsymbol{x}_0;\boldsymbol{x})\}\,\mathrm{d}S(\boldsymbol{x})\\
&\quad \quad +\llbracket {E}_{n}(\boldsymbol{x}_0)\rrbracket \left[\frac{Q}{Q-1}-L(\boldsymbol{x}_0)  \right]=E_{n0}+\frac{q(\boldsymbol{x}_0)}{Q-1}, \qquad \mbox{for}\,\,\,\boldsymbol{x}_{0}\in S.
\label{eq:intjump}
\end{split}
\end{align}
The jump $\llbracket {E}_{n}\rrbracket$ can therefore be computed from \eqref{eq:intjump} for a given surface charge density after discretization of the integral  on a mesh, yielding a large linear system that is solved iteratively. Further details of the numerical implementation are given in \S \ref{sec:numerical} and in appendix~A. Having obtained $\llbracket {E}_{n}\rrbracket$, the normal components $E_n^+$ and $E_n^-$ are easily obtained using equation (\ref{eq:normalelectric}).

The tangential component of the electric field can then be evaluated using \eqref{eq:intelectric}; however, care must be taken to remove the strong singularity in the kernel. Here, we adopt instead an indirect method in which we first compute the electric potential $\varphi$ using equation \eqref{eq:intpotential} then differentiate it numerically on the drop surface to obtain $\boldsymbol{E}_t$. Once the normal and tangential components of the electric field are known, we can determine the jump in the normal component of Ohmic currents $\llbracket \sigma E_n\rrbracket$ as well as the jump in electric tractions $\llbracket \boldsymbol{f}^E\rrbracket $ using equation \eqref{eq:electraction}.

\subsection{Flow problem}\label{sec:flow}

The applied electric field induces fluid motion inside and outside the drop. The need to solve for the fluid flow is twofold, as it affects the surface charge distribution according to equation \eqref{eq:chargeeq0} and causes deformations of the interface, which is a material surface advected by the flow. The flow problem is solved after application of the dynamic boundary condition \eqref{eq:stressbalance} to obtain the hydrodynamic traction jump $\llbracket \boldsymbol{f}^H\rrbracket $ on the drop-fluid interface. Assuming creeping flow, we use the Stokes boundary integral equation to represent  the fluid velocity as \citep{rallison1978,pozrikidis2002}
\begin{align}
\begin{split}
\boldsymbol{v}(\boldsymbol{x}_0)=&-\frac{1}{4\uppi \mu (1+\lambda)} \oint_S \llbracket \boldsymbol{f}^H(\boldsymbol{x})\rrbracket  \boldsymbol{\cdot} \mathsfbi{G} (\boldsymbol{x}_0;\boldsymbol{x}) \,\mathrm{d}S(\boldsymbol{x})\\
&+\frac{\kappa}{4 \uppi} \oint_S \boldsymbol{v}(\boldsymbol{x}) \boldsymbol{\cdot} \mathsfbi{T}(\boldsymbol{x}_0;\boldsymbol{x})\boldsymbol{\cdot} \boldsymbol{n}(\boldsymbol{x}) \,\mathrm{d}S(\boldsymbol{x}), \quad \mbox{for}\,\,\, \boldsymbol{x}_0 \in V^\pm, S, \label{eq:stokesbie}
\end{split}
\end{align}
where $\kappa=(1-\lambda)/(1+\lambda)$ and $\mathsfbi{G}(\boldsymbol{x}_0;\boldsymbol{x})$ and $\mathsfbi{T}(\boldsymbol{x}_0;\boldsymbol{x})$ denote the free-space Green's functions for the Stokeslet and stresslet, respectively:
\refstepcounter{equation}
$$
\mathsfbi{G}(\boldsymbol{x}_0;\boldsymbol{x})=\frac{\mathsfbi{I}}{r} + \frac{\boldsymbol{r}\boldsymbol{r}}{r^3},  \qquad \mathsfbi{T}(\boldsymbol{x}_0;\boldsymbol{x})=6\frac{\boldsymbol{r}\boldsymbol{r}\boldsymbol{r}}{r^5}. \eqno{(\theequation{\mathit{a},\mathit{b}})} 
 \label{eq:stokeslet}
$$
The usual negative sign in the definition of the stresslet appears if $\boldsymbol{r}$ is defined as $\boldsymbol{x}-\boldsymbol{x}_0$. Note that $\kappa=\pm 1$ corresponds to the case of a bubble ($\lambda \rightarrow 0$) and solid particle ($\lambda \rightarrow \infty$), respectively. The interfacial velocity appearing in the double layer potential is yet unknown, but an integral equation for $\boldsymbol{v}$ on the surface can be obtained by moving the evaluation point $\boldsymbol{x}_0$ to the boundary $S$. In dimensionless form, it reads:
\begin{align}
\begin{split}
\boldsymbol{v}(\boldsymbol{x}_0)+&\frac{\lambda-1}{8\upi} \oint_S [\boldsymbol{v}(\boldsymbol{x}) - \boldsymbol{v}(\boldsymbol{x}_0)] \boldsymbol{\cdot} \mathsfbi{T}(\boldsymbol{x}_0;\boldsymbol{x})\boldsymbol{\cdot} \boldsymbol{n}(\boldsymbol{x}) \,\mathrm{d}S(\boldsymbol{x})\\
=&-\frac{1}{8\uppi Ma} \oint_S \llbracket \boldsymbol{f}^H(\boldsymbol{x})\rrbracket  \boldsymbol{\cdot} \mathsfbi{G} (\boldsymbol{x}_0;\boldsymbol{x}) \,\mathrm{d}S(\boldsymbol{x}), \qquad \mbox{for}\,\,\, \boldsymbol{x}_0 \in S. \label{eq:stokessurface}
\end{split}
\end{align}
The forcing term in this equation is contained in the hydrodynamic traction jump $\llbracket \boldsymbol{f}^H \rrbracket$. After discretization of  the integral, equation \eqref{eq:stokessurface} yields a dense linear system that is again solved iteratively. The weak singularity appearing in the double-layer potential in the original equation \eqref{eq:stokesbie} has been removed by using appropriate integral identities; the weak singularity of the single-layer potential, on the other hand, disappears after introducing plane polar coordinates \citep{pozrikidis1992}. It is well known that the integral equation \eqref{eq:stokessurface} admits arbitrary rigid body motions and uniform expansion as eigensolutions, resulting in the ill-conditioning of the linear system for $\lambda \gg 1 $ or $\lambda \ll 1$ and leading to poor convergence of the solution \citep{zinchenko1997}. We employ Wielandt's deflation technique to eliminate $\kappa=\pm 1$ from the spectrum of the integral equation to cure the ill-conditioning \citep{kim2013}; see appendix B for details. Once the interfacial velocity is known, the nodes are advected with the normal component of the fluid velocity; the heuristic mesh relaxation algorithm of \cite{loewenberg1996} is applied in the tangential direction so as to reduce mesh distortion.

\subsection{Summary of the numerical method} \label{sec:numerical}

\begin{figure}
\centering
\includegraphics[width=0.9\linewidth]{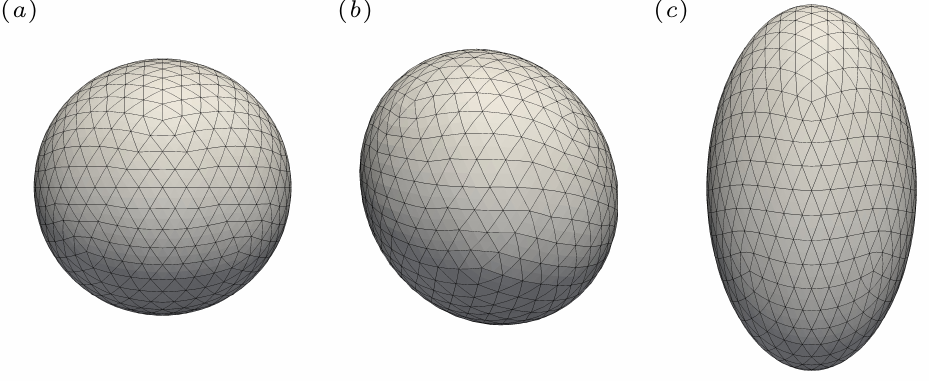}
\caption{Discretized mesh: $N_\triangle = 1280$ six-node curved elements. (\textit{a}) An initially spherical mesh at time $t=0$, (\textit{b}) a deformed mesh for a tilted drop in the Quincke regime corresponding to the case of figure \ref{fig:snapquincke}, and (\textit{c}) a deformed mesh of a prolate drop in the Taylor regime (system 3), where we applied the mesh relaxation algorithm of \citet{loewenberg1996}.}
\label{fig:mesh}
\end{figure}

We solve integral equations \eqref{eq:intpotential}, \eqref{eq:intjump} and \eqref{eq:stokessurface} numerically using the boundary element method on a discrete representation of the drop surface \citep{pozrikidis2002}. The initially spherical surface is first discretized by successive subdivision of an icosahedron, by which each triangular element is subdivided into four new triangles whose nodes are projected onto the sphere \citep{loewenberg1996}. This leads to a highly uniform triangular mesh, in which we treat each element as a six-node curved element thus allowing for computation of the local curvature. Most of the results we present here are on a surface with $N_\triangle =320$ elements and 642 nodes obtained after $N_d=2$ successive subdivisions, though a few results are also shown with $N_\triangle =1280$ elements and 2562 nodes, corresponding to $N_d=3$ subdivisions. Typical meshes with $N_d=3$ are shown in figure~\ref{fig:mesh} for different levels of deformation. The evaluation of integrals and the calculation of geometrical properties such as the unit normal and curvature on the discretized surface are  standard and are outlined in appendix~A. 

The numerical algorithm during one integration step can be summarized as follows:
\begin{itemize}
\item Given an interfacial charge distribution $q$ (which is taken to be uniformly zero at $t=0$), solve for $\llbracket E_n\rrbracket$, $E_n^+$ and $E_n^-$ by inverting equation (\ref{eq:intjump}) numerically, together with equation (\ref{eq:normalelectric}). Discretization of the integrals in (\ref{eq:intjump}) yields a large algebraic system which we solve iteratively using GMRES \citep{saad1986}. 
\item Evaluate the electric potential $\varphi$ on the drop surface using equation (\ref{eq:intpotential}), where the single-layer density $\llbracket E_n\rrbracket$ is known. 
\item Differentiate $\varphi$ on the drop surface using the method outlined in appendix~A to obtain the tangential component $\boldsymbol{E}_{t}=-(\mathsfbi{I}-\boldsymbol{nn})\bcdot\bnabla \varphi$ of the electric field. 
\item Calculate the jump in hydrodynamic tractions $\llbracket \boldsymbol{f}^{H}\rrbracket$ using the dynamic boundary condition (\ref{eq:stressbalance}), where electric tractions and surface tension forces are known from the solution of the electric problem and from the current geometry.
\item Solve for the interfacial velocity $\boldsymbol{v}$ by inverting the boundary integral equation (\ref{eq:stokessurface}), which again yields an algebraic system after discretization of the integrals. 
\item Update the surface charge density $q$ and advance the position of the surface nodes $\boldsymbol{x}_{i}$ by numerical integration of the charge conservation equation and kinematic boundary condition,
\begin{align}
&\frac{\partial q}{\partial t}= \frac{Q+2}{1+2R}(E_n^--RE_n^+)- \boldsymbol{\nabla}_s\boldsymbol{\cdot}(q\boldsymbol{v}_t)+ \boldsymbol{v}_m \boldsymbol{\cdot}\boldsymbol{\nabla}_s q, \label{eq:chargeeq} \\
&\frac{\mathrm{d} \boldsymbol{x}_{i}}{\mathrm{d}t} = \boldsymbol{n}(\boldsymbol{x}_{i})\boldsymbol{\cdot}\boldsymbol{v}(\boldsymbol{x}_{i})+ \boldsymbol{v}_m(\boldsymbol{x}_{i}),
\label{eq:nodeadvect}
\end{align}
where $\boldsymbol{v}_{m}$ denotes the tangential mesh relaxation velocity and is determined using the method proposed by \cite{loewenberg1996}. Numerical integration of these equations is performed explicitly in time using a second-order Runge-Kutta scheme. 
\end{itemize}

The charge conservation equation (\ref{eq:chargeeq}) requires numerical evaluation of the surface divergence and gradient appearing on the right-hand side. These quantities are obtained by analytical differentiation based on the parametrization discussed in appendix~A; an alternate method based on finite volumes \citep{yon1998}, and a semi-implicit scheme wherein the linear $\llbracket \sigma E_n\rrbracket$ and nonlinear $\boldsymbol{\nabla}_s\boldsymbol{\cdot}(q\boldsymbol{v})$ terms are treated implicitly and explicitly, respectively, were also attempted but did not produce significant differences in the results. The numerical method was tested extensively by first considering the case of a solid spherical particle under Quincke rotation, for which an exact analytical solution based on spherical harmonics is available \citep{das2013}, and by comparison with previous numerical studies of drop dynamics in simple shear flow \citep{kennedy1994} and under electric fields in the absence of charge convection \citep{lac2007}.

\section{Results and discussion} \label{sec:results}

We now turn to simulation results, which we compare with existing experimental data. Following prior studies, we characterize deviations from the spherical shape using Taylor's deformation parameter $\mathcal{D}$, which we define as 
\begin{equation}
\mathcal{D}=\frac{L-B}{L+B}.
\end{equation}
In axisymmetric configurations (Taylor regime), $L$ and $B$ denote the lengths of the drop axes in directions parallel and perpendicular to the electric field, respectively, so that the sign of $\mathcal{D}$ distinguishes between  oblate ($\mathcal{D}<0$) and prolate ($\mathcal{D}>0$) shapes. When electrorotation takes places (Quincke regime), $L$ and $B$ are defined as in figure~\ref{fig:figure1} as the lengths of the major and minor axes of the drop, respectively, so that $\mathcal{D}>0$ at all times. We also introduce the tilt angle $\alpha$ as the angle between the major axis of the drop and the plane normal to the applied field, where $\alpha=0$ in the Taylor regime and $\alpha>0$ in the Quincke regime. The determination of these geometric quantities is performed by fitting an ellipsoid to the drop surface using a least-squares algorithm.

\subsection{Taylor regime}\label{sec:taylorregime}

\begin{table}
\vspace{-0.3cm}
  \begin{center}
  \begin{tabular}{cccccccccc}
     System & $\epsilon^+/\epsilon_0$ & $\epsilon^-/\epsilon_0$ & $\sigma^+$ & $\sigma^-$ & $\mu^+$ & $\mu^-$ & $\gamma$  & $a$ &$E_0$ \\[1pt]
      & & & (S $\text{m}^{-1}$) & (S $\text{m}^{-1}$)& (Pa s)& (Pa s) & (mN $\text{m}^{-1}$) & (mm) & (kV $\text{cm}^{-1}$) \\  [5pt]
     1a & 4.9 & 2.8 & $5.8 \times 10^{-11}$  & $0.2 \times 10^{-11}$ & 0.68 & 0.05 & 4.5 & 2.0 & 1.6 \\[0pt]
     1b & 4.9 & 2.8 & $5.8 \times 10^{-11}$  & $0.2 \times 10^{-11}$ & 0.68 & 0.05 & 4.5 & 2.0 & 2.1 \\[0pt]
     1c & 4.9 & 2.8 & $5.8 \times 10^{-11}$  & $0.2 \times 10^{-11}$ & 0.68 & 0.05 & 4.5 & 2.0 & 6.1 \\[0pt]
     2a & 5.3 & 3.0 & $4.5 \times 10^{-11}$  & $0.12 \times 10^{-11}$ & 0.69 & 0.97 & 4.5 & 0.7 & 0.45--2.0 \\[0pt]
     2b & 5.3 & 3.0 & $4.5 \times 10^{-11}$  & $0.12 \times 10^{-11}$ & 0.69 & 0.97 & 4.5 & 2.1 & 0.26--1.2 \\ \hline 
  \end{tabular}
  \caption{Material properties: systems 1 and  2 correspond to the experiments of \citet{lanauze2015} and \citet{salipante2010}, respectively. $\epsilon_0=8.8542 \times 10^{-12}\,\text{F.m}^{-1}$ denotes the permittivity of vacuum.} \label{table:dimensionaltaylor}
  \end{center}
\end{table} 

\begin{table}
  \begin{center}
\def~{\hphantom{0}}
  \begin{tabular}{ccccccc}
     System & $R$ & $Q$ & $\lambda$ & $Ca_E$ & $Ma$ \\ [5pt]
     1a & 29.0 & 0.57 & 0.074 & 0.49 & 0.65 \\[0pt]
     1b & 29.0 & 0.57 & 0.074 & 0.85 & 0.375 \\[0pt]
     1c & 29.0 & 0.57 & 0.074 & 7.18 & 0.045 \\[0pt]
     2a & 36.6 & 0.57 & 1.41 & 0.03--0.6 & 0.27--5.4 \\[0pt]
     2b & 36.6 & 0.57 & 1.41 & 0.03--0.6 & 0.8--16\\[0pt]
     3 & 0.1 & 1.37 & 1 & 0.3 & 0.5 \\ \hline
  \end{tabular}
  \caption{Dimensionless parameters corresponding to the material properties of table 1:  systems 1, 2 and 3 correspond to the experiments of \citet{lanauze2015}, \citet{salipante2010} and \citet{ha2000a}, respectively.}
  \label{table:dimensionlesstaylor}
  \end{center}
\end{table}

We first investigate drop dynamics in the Taylor regime, where the drops attain either a steady oblate or prolate shape depending on material properties. The Taylor regime was  addressed in our recent work using both a small-deformation theory and axisymmetric boundary element simulations \citep{das2016}, and is primarily used here as a benchmark for our three-dimensional algorithm. Material properties in our simulations are chosen based on the experiments of \citet{lanauze2015} for transient (system 1) and \citet{salipante2010} for steady drop deformations (system 2) and are provided in table~\ref{table:dimensionaltaylor}; corresponding dimensionless parameters are presented in table~\ref{table:dimensionlesstaylor}. Both of these experiments focused on oblate drops. We also consider the case of prolate deformations using one set of parameters from the experiments of \citet{ha2000a} (system 3); their study, however, did not report all the material properties necessary to construct the five dimensional groups required in our model, so we arbitrarily set the electric capillary number and Mason number values to $Ca_E=0.3$ and $Ma=0.5$, respectively.

\begin{figure}
\centering
\includegraphics[width=\linewidth]{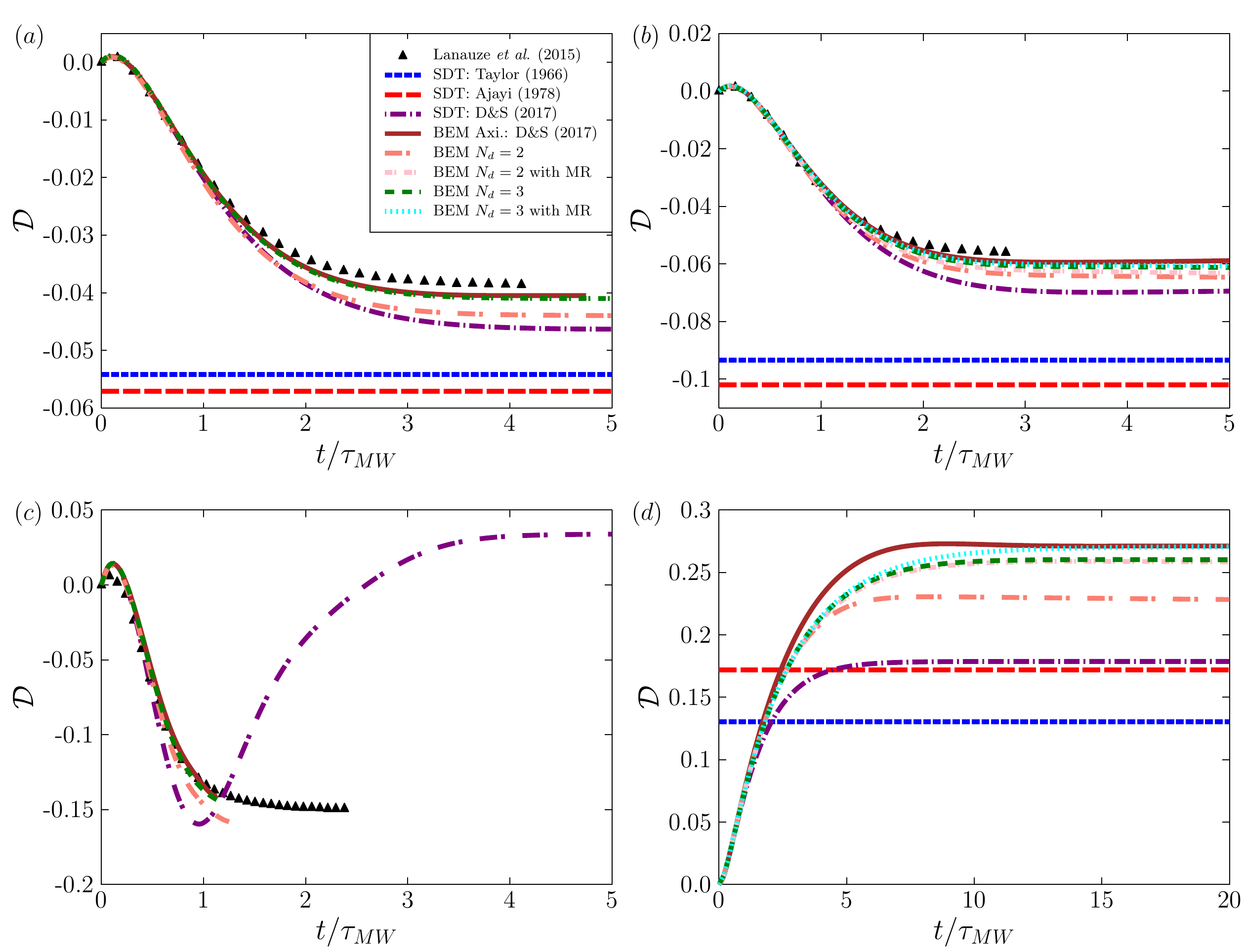}
\caption{(Color online) Deformation parameter $\mathcal{D}$ as a function of time for the parameters of: ($a$) system 1a, ($b$) system 1b,  ($c$) system 1c, and ($d$) system 3. Boundary element (BEM) results are compared to the experiments of \cite{lanauze2015} in the case of oblate drops, and to various small-deformation theories (SDT). The steady deformation values predicted by the models of Taylor (1966) and Ajayi (1978) in the case of system 1c are $-0.75$ and $-1.40$, respectively, and out of the frame of the figure. The effect of the mesh relaxation (MR) algorithm is also shown and found to be greater when large deformations arise (system 3).}
\label{fig:transienttaylor}
\end{figure}

Figure \ref{fig:transienttaylor}($a$) shows the transient deformation of an oblate drop corresponding to system 1a for an electric field strength of $E_0/E_c=0.49$. Unsurprisingly, the axisymmetric boundary element method performs best in predicting the drop deformation when compared with experiments. Results from our three-dimensional simulations are shown for two different mesh resolutions ($N_d=2$ and $3$) as a convergence test; we find as expected that the accuracy improves with increasing $N_d$, and the results with $N_d=3$ are nearly identical to the predictions of the axisymmetric code. The classic small-deformation theories of \cite{taylor1966} and \cite{ajayi1978}  that neglect interfacial charge convection perform rather poorly; however, inclusion of charge convection in the theoretical model improves the results considerably \citep{das2016}. 

\begin{figure}
\centering
\includegraphics[width=0.95\linewidth]{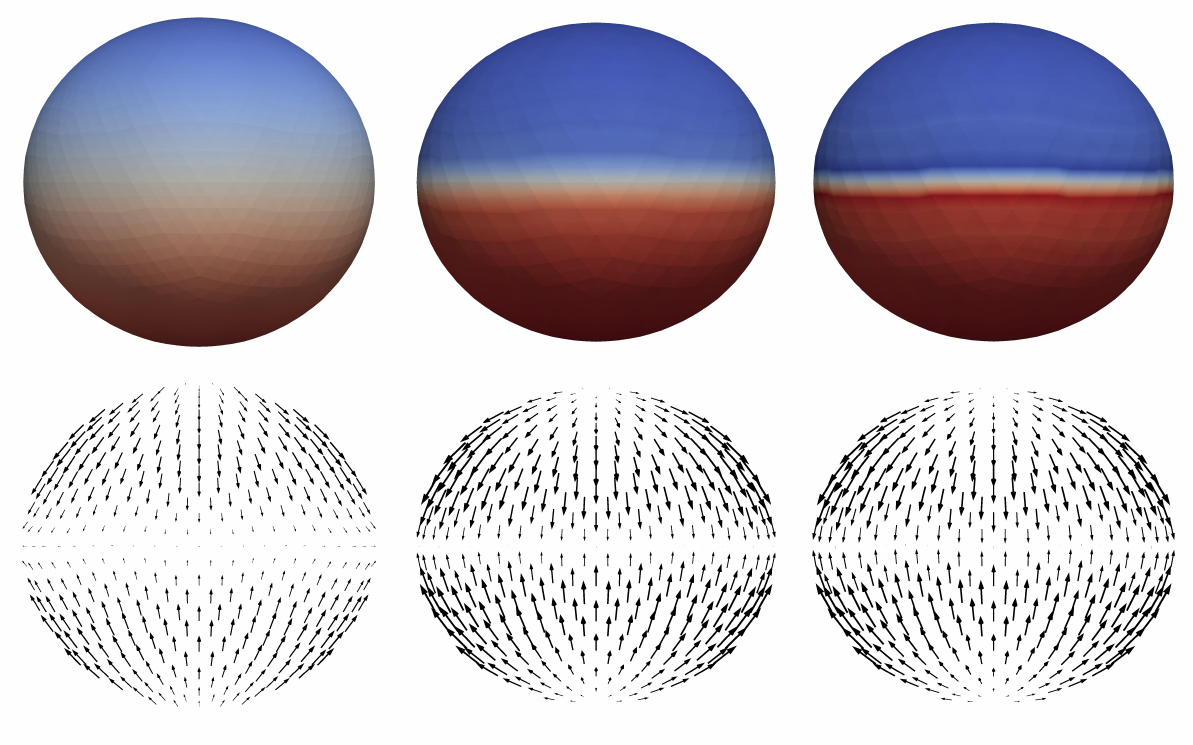}
\caption{(Color online) Time evolution profiles of the surface charge density (top row) and interfacial fluid velocity (bottom row) in the case of system 1b in the Taylor regime at $t/\tau_{MW}=1.0$, $2.5$ and  $4.0$ (left to right). See supplementary online materials for a movie showing the dynamics and flow field in this case.}
\label{fig:snaptaylor}
\end{figure}

The case of system 1b, corresponding to a stronger applied field ($E_0/E_c=0.64$), shows the same trends albeit with larger deformations in figure~\ref{fig:transienttaylor}($b$). While the boundary element simulations capture the transient and steady-state accurately, the performance of small-deformation theories is not as good as previously due to significant deformations. The surface charge distribution and fluid velocity obtained from the three-dimensional simulation for this case are illustrated at three different times in figure~\ref{fig:snaptaylor}. As revealed by these snapshots, the interfacial velocity, which is directed from the poles towards the equator, causes transport of negative and positive charges towards the equatorial circumference of the drop, thereby inducing a sharp charge gradient across it. This gradient cannot be captured by small-deformation theories, as these employ truncated spherical harmonic expansions to represent variables; it is also challenging to capture numerically, especially as $E_0/E_c$ is increased further.    

This is illustrated in figure~\ref{fig:transienttaylor}($c$), showing the case of system 1c with an even higher electric field of $E_0/E_c=1.86$. There, the charge gradient across the interface becomes sharper and an actual discontinuity appears that triggers instabilities, reminiscent of Gibbs phenomenon, leading to the termination of the simulations. \citet{lanauze2015} were the first to discover this charge shock in their numerical work, and suggested that it might be an artefact of the axisymmetric nature of their boundary element simulations, which prevents transition to Quincke electrorotation. As we demonstrate here, the development of the charge shock in fact can occur in the Taylor regime, where it is  due to the quadrupolar Taylor flow in the case of oblate drops that causes the sweeping of positive and negative charges towards the equator. The strength of this flow increases with electric field and is more pronounced for low-viscosity drops, leading to stronger shocks in these cases. While more analysis is required to understand the detailed structure of these shocks, we note that the Melcher--Taylor leaky dielectric model does not account for charge diffusion, which may have a regularizing effect in experiments. As expected, figure~\ref{fig:transienttaylor}($c$) shows a very poor performance of small-deformation theories in this regime, which are slightly improved by inclusion of charge convection but are unable to capture the charge discontinuity. 

The case of prolate drop deformations corresponding to system 3 is shown in figure~\ref{fig:transienttaylor}($d$), where larger deformations are observed.  The steady state deformation value reported in the experiments of \citet{ha2000a}, which did not specify the value of $Ma$, is $\mathcal{D}=0.25$; the simulations of \citet{lac2007} with $Ma \rightarrow \infty$ reported $\mathcal{D}=0.22$, while our  simulations with $Ma=0.5$ predict $\mathcal{D}=0.27$. No experimental data exist for the transient deformation, so we use axisymmetric simulations as the benchmark in this case. We find as expected that the three-dimensional simulations with $N_d=3$ perform best, especially when the mesh relaxation algorithm is used as deformations are significant. Unsurprisingly, the large drop deformation is poorly captured and underpredicted by the various small-deformation theories.

\begin{figure}
\centering
\includegraphics[width=\linewidth]{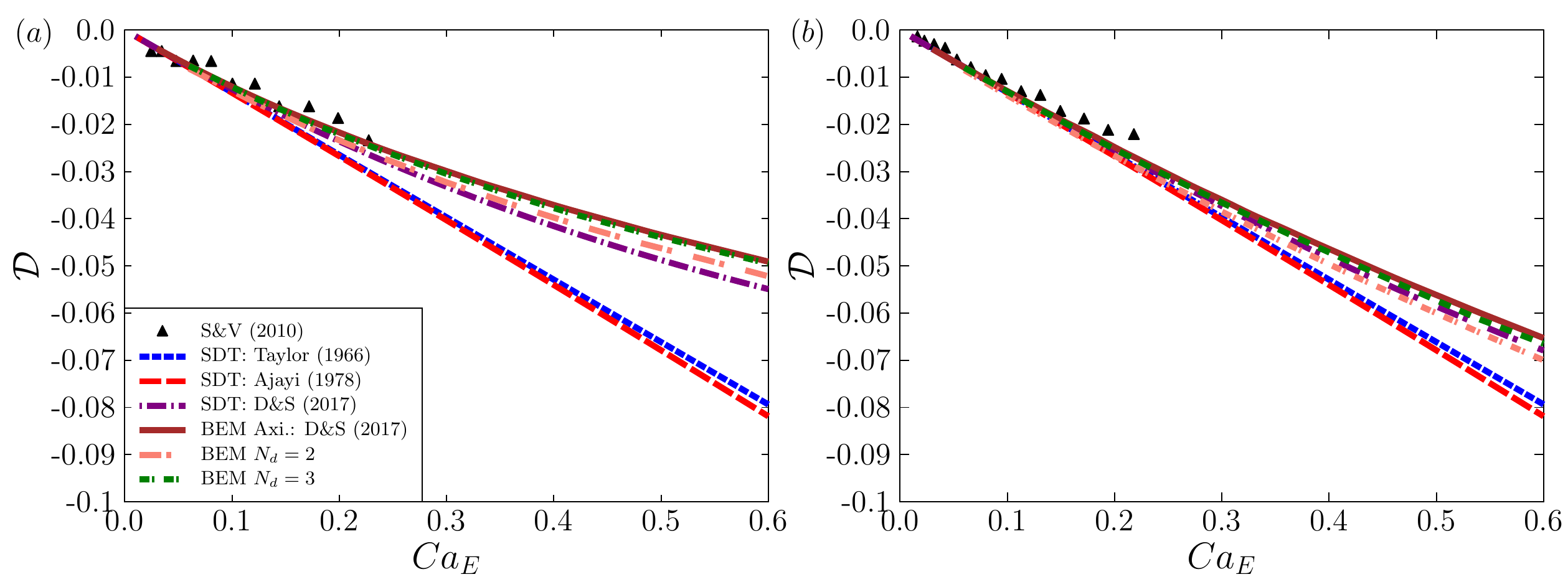}
\caption{(Color online) Steady drop deformation $\mathcal{D}$ as a function of electric capillary number $Ca_E$ for the parameters of: ($a$) system 2a, and ($b$) system 2b. Boundary element (BEM) results are compared to the experiments of \citet{salipante2010} and to various small-deformation theories (SDT).}
\label{fig:steadytaylor}
\end{figure}

We conclude the discussion of the Taylor regime by considering steady state drop deformations corresponding to system 2, for which we compare our simulations with theoretical and experimental data in figure~\ref{fig:steadytaylor}. Steady deformation values are shown for increasing values of electric capillary number $Ca_E$ for two different drop sizes of $a=0.7\,\mathrm{mm}$ and $a=2.1\,\mathrm{mm}$. For a given value of $Ca_E$, the smaller drop experiences a stronger electric field corresponding to a lower value of $Ma$ when compared to the larger drop. As a consequence, the small drop experiences stronger charge convection on its surface, which tends to reduce deformations as previously shown by other authors \citep{feng1999,lanauze2015}. In consistency with previous results, the axisymmetric and three-dimensional simulations perform best followed by the small-deformation theory with convection \citep{das2016}. Since the effect of convection is weaker in the case of the larger drop, the small-deformation theories without convection do not deviate as much from the experimental data and simulation results as for the smaller drop.

\subsection{Quincke regime}

We now turn our attention to the electrorotation of drops in the Quincke regime, which is seen to occur when the applied field exceeds a certain critical value. For comparison with experiments, we use the parameter values provided by \citet{salipante2010} but restrict ourselves to small drop sizes. We consider two different sets of material properties which are summarized in tables~4 and 5 and correspond to different viscosity ratios. The heuristic mesh relaxation algorithm of \citet{loewenberg1996} is not included in the simulations in the Quincke regime, as we found that it caused numerical instabilities preventing the simulations from reaching steady state; as deformations tend to be fairly moderate when electrorotation takes place ($\mathcal{D} \lesssim 0.1$ in the simulations shown below), we do not expect significant errors due to mesh distortion. 

\begin{table}
  \begin{center}
\def~{\hphantom{0}}
  \begin{tabular}{cccccccccc}
     System & $\mu^+$ & $\mu^-$ & $\gamma$  & $a$ &$E_0$ \\[0pt]
       & (Pa.s)& (Pa.s) & (mN.$\text{m}^{-1}$) & (mm) & (kV.$\text{cm}^{-1}$) \\  [5pt]
     2c & 0.69 & 9.74 & 4.5 & 0.25, 0.75, 1.25, 1.75 & 0.67--5.36 \\[0pt]
     2d & 0.69 & 4.87 & 4.5 & 0.25, 0.75, 1.25, 1.75 & 0.67--5.36 \\[0pt]
 \hline 
  \end{tabular}
  \caption{Material properties for system 2, corresponding to the experiments of \citet{salipante2010} with a critical electric field of $E_c=2.68$ kV.$\text{cm}^{-1}$. The permittivity and conductivity values for this system are given in table \ref{table:dimensionaltaylor}.}
  \label{table:dimensionalquincke}
  \end{center}
\end{table}

\begin{table}
  \begin{center}
\def~{\hphantom{0}}
  \begin{tabular}{ccccccc}
     System & $R$ & $Q$ & $\lambda$ & $Ca_{MW}$ & $E_0/E_c$ \\ [5pt]
     2c & 36.6 & 0.57 & 14.1 & 0.44, 1.32, 2.20, 3.08 & 0.25--2.0 \\[0pt]
     2d & 36.6 & 0.57 & 7.05 & 0.23, 0.69, 1.15, 1.61 & 0.25--2.0 \\[0pt]
     \hline
  \end{tabular}
  \caption{Dimensionless parameters corresponding to the material properties shown in table \ref{table:dimensionalquincke} for system 2, obtained from the experiments of \citet{salipante2010}.}
  \label{table:dimensionlessquincke}
  \end{center}
\end{table}

\begin{figure}
\centering
\includegraphics[width=0.95\linewidth]{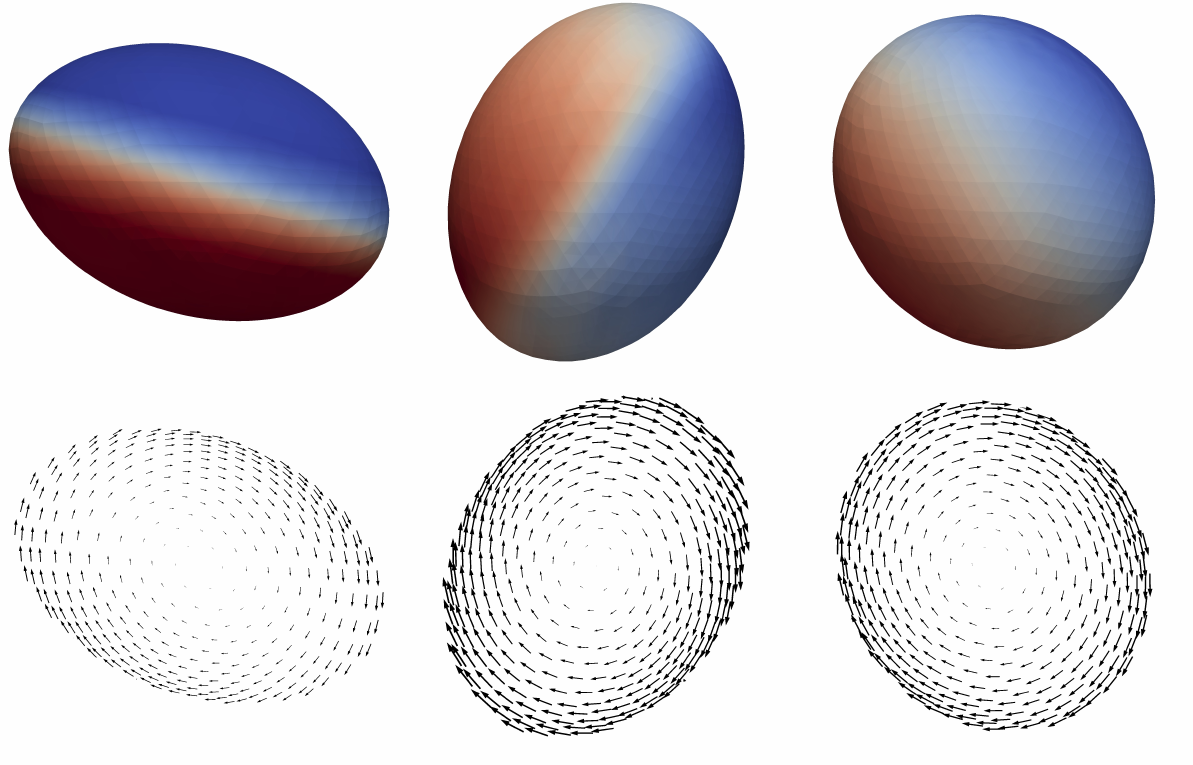}
\caption{(Color online) Time evolution profiles of the surface charge density (top row) and interfacial fluid velocity (bottom row) in the case system 2c in the Quincke regime at $t/\tau_{MW}=3.75$, $5.25$ and  $10.5$ (left to right). See supplementary online materials for a movie showing the dynamics and flow field in this case.}
\label{fig:snapquincke}
\end{figure}

\begin{figure}
\centering
\includegraphics[width=\linewidth]{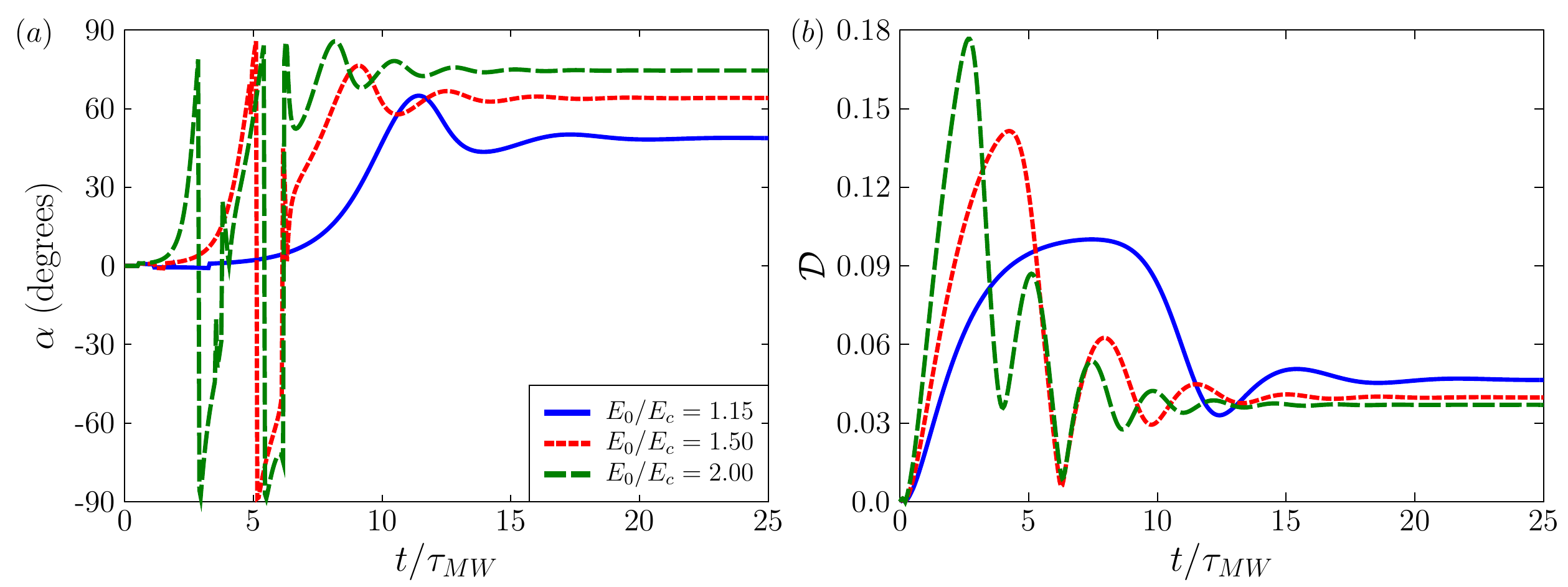}
\caption{(Color online) (\textit{a}) Tilt angle $\alpha$ and (\textit{b}) drop deformation parameter $\mathcal{D}$ as functions of time $t/\tau_{MW}$ for system 2d with drop size $a=0.75$ mm and $Ca_{MW}=0.69$. Stronger electric fields cause faster and more pronounced oscillations in the tilt angle and drop deformation.}
\label{fig:transientquincke}
\end{figure}

A typical simulation exhibiting Quincke rotation is illustrated in figure~\ref{fig:snapquincke} in the case of system 2c for an initial drop radius of  $a=1.25\,\mathrm{mm}$ and electric field $E_0/E_c=1.5$, where $E_c$ is the critical electric field for the onset of rotation of a rigid sphere given in equation~(\ref{eq:Ec}). The figure shows both the interfacial charge profile and interfacial velocity field at different times during the transient. Upon application of the field, the drop deforms towards an oblate shape similar to that found in the Taylor regime. This configuration, however, becomes unstable and leads to the rotation of the drop with respect to an arbitrary axis perpendicular to the field direction. As it rotates, the drop  relaxes towards a more spherical shape as we characterize in more detail below, and ultimately reaches a steady shape with a tilt angle $\alpha$ with respect to the horizontal plane. As is visible in figure~\ref{fig:snapquincke}, the charge profile is smoother than in the Taylor regime and is no longer axisymmetric, leading to a net electrostatic dipole that forms an angle with the field direction; the nature of the flow is also significantly different from the classic Taylor flow and appears to be primarily rotational. The transient dynamics are illustrated in more detail in figure~\ref{fig:transientquincke}, showing the tilt angle $\alpha$ and deformation parameter $\mathcal{D}$ as functions of time for different electric field strengths. Oscillations in both $\alpha$ and $\mathcal{D}$ are observed during the transient and are more significant in stronger fields, where the drop can undergo actual tumbling before its orientation stabilizes; similar time dynamics have also been reported in experiments \citep{salipante2010} and theory \citep{he2013}. In yet stronger fields, experiments have shown that the dynamics in some cases do not reach a steady state but instead exhibit chaotic tumbling and stretching of the drop \citep{salipante2013}; this regime was not captured in our simulations, which became unstable in very strong fields.  

\begin{figure}
\centering
\includegraphics[width=\linewidth]{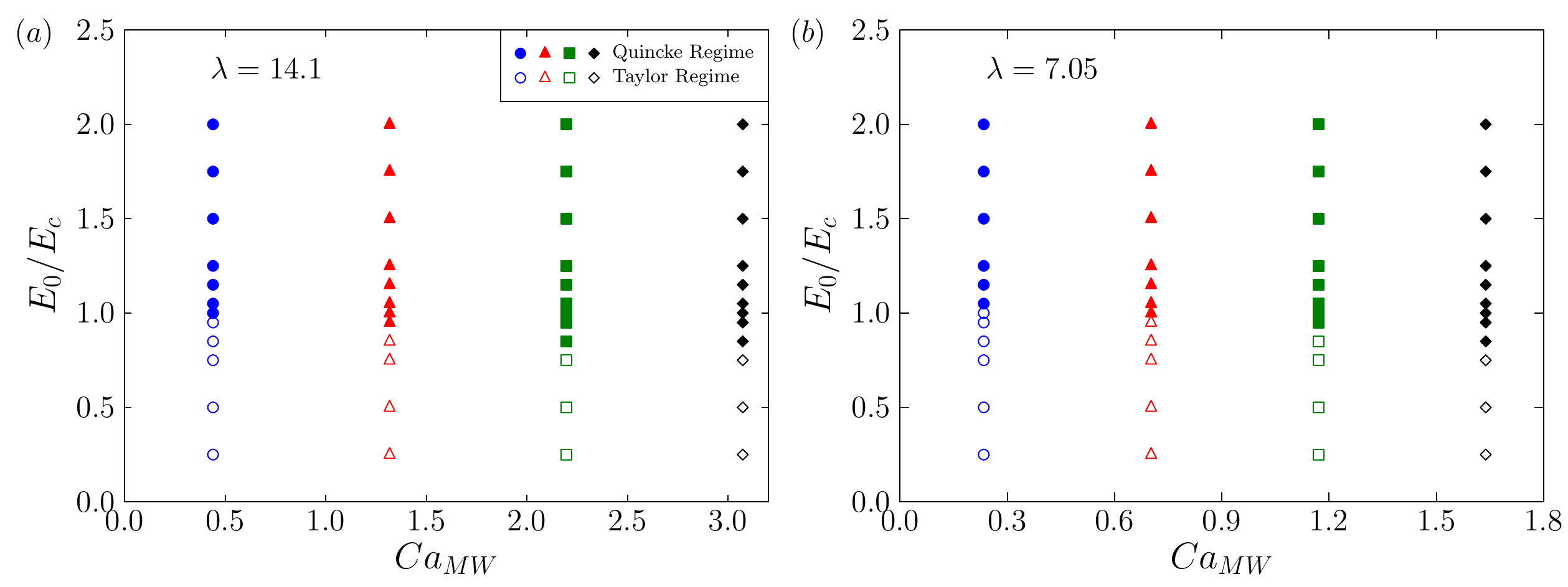}
\caption{(Color online) Phase diagram distinguishing the axisymmetric Taylor regime (empty symbols) from the Quincke electrorotation regime (filled symbols) for two different viscosity ratios: (\textit{a}) $\lambda=14.1$, and (\textit{b}) $\lambda=7.05$.}
\label{fig:phasequincke}
\end{figure}

The transition from the Taylor regime to the Quincke regime is characterized in more detail in figure \ref{fig:phasequincke} showing phase diagrams for systems 2c and 2d in the $(E_0/E_c,Ca_{MW})$ plane, where we recall that for fixed material properties $Ca_{MW}$ is a measure of drop size. The case of a very viscous drop ($\lambda = 14.1$) is shown in figure \ref{fig:phasequincke}($a$), where the critical electric field for the transition to electrorotation is found to be close to the value of $E_c$ for a rigid sphere, yet decreases slightly with increasing $Ca_{MW}$. A small highly viscous drop is indeed expected to behave in the same way as a rigid particle. Increasing $Ca_{MW}$ (or equivalently, drop size) at a fixed value of $E_0/E_c$ leads to larger deformations in the Taylor regime, which causes an increase in the effective dipole induced inside the drop and thus has a destabilizing effect as demonstrated by the decrease in the critical electric field. A similar phase diagram is obtained at the lower viscosity ratio of $\lambda=7.05$ in figure \ref{fig:phasequincke}($b$); decreasing $\lambda$, however, is found to slightly increase the threshold for Quincke rotation. All of these trends are consistent with the experimental data of \cite{salipante2010}. 

\begin{figure}
\centering
\includegraphics[width=\linewidth]{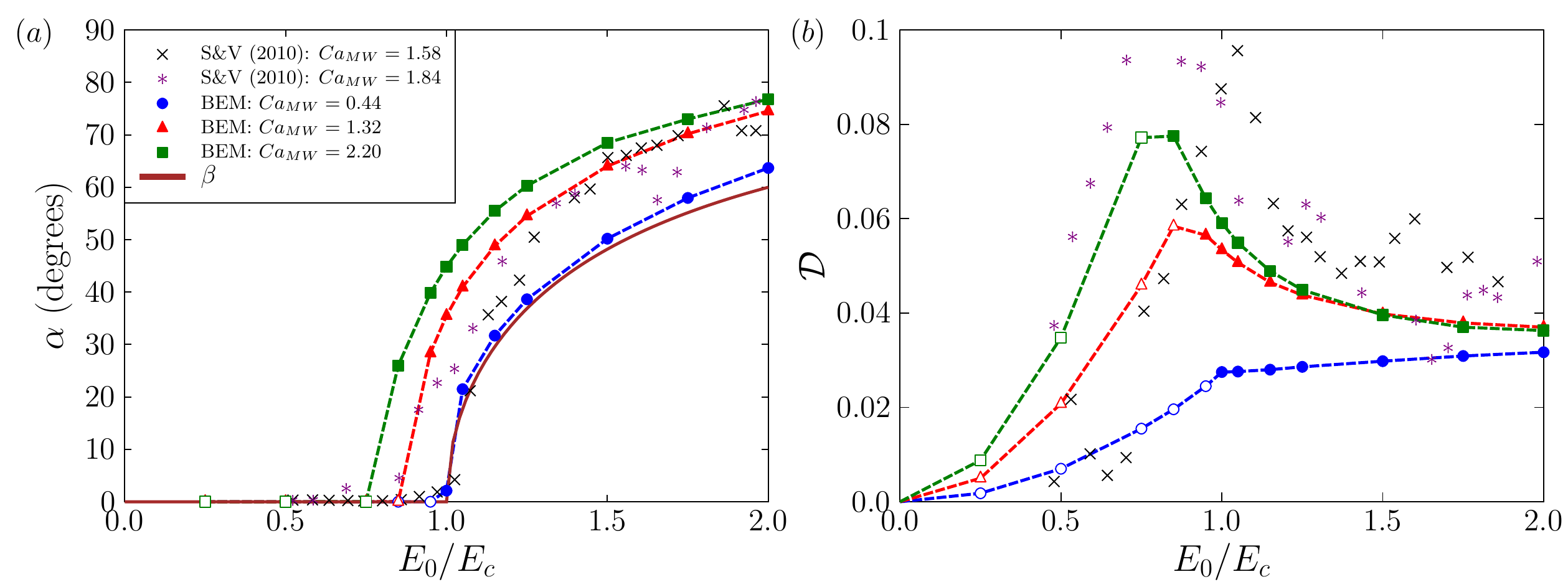}
\caption{(Color online) (\textit{a}) Steady tilt angle $\alpha$ and (\textit{b}) drop deformation parameter $\mathcal{D}$ as functions of applied electric field strength $E/E_c$ for system 2c for different values of $Ca_{MW}$. Boundary element (BEM) simulation results are compared with the experiments of \citet{salipante2010}.}
\label{fig:tiltlambda14}
\end{figure}

The steady-state tilt angle $\alpha$ is shown as a function of electric field strength in figure~\ref{fig:tiltlambda14}($a$) for system 2c, where  it is also compared with the complementary of the angle between the steady dipole and applied electric field in the case of a rigid sphere, which we denote by $\beta$ \citep{salipante2010}:
\begin{equation}
\beta = \frac{\uppi}{2} - \arctan{\left[ \left(\frac{E_0^2}{E_c^2}-1\right)^{-1/2} \right]}.
\end{equation}
In the Taylor regime, the tilt angle is zero as the drop shape is axisymmetric. As field strength increases, a supercritical pitchfork bifurcation is observed at the onset of rotation, with a value of $\alpha$ that increases with $E_0/E_c$ and asymptotes towards $\upi/2$ in strong fields. Both angles $\alpha$ and $\beta$ show similar trends as expected, especially in the case of weakly deformed drops ($Ca_{MW}=0.44$) that behave like rigid spheres. Increasing drop size (or equivalently $Ca_{MW}$) causes the bifurcation to occur at lower field strengths in agreement with the phase diagram of figure~\ref{fig:phasequincke}. These trends once again agree with the experimental results of \citet{salipante2010} at similar values of $Ca_{MW}$. 

Corresponding values of the steady drop deformation $\mathcal{D}$ are shown in figure~\ref{fig:tiltlambda14}($b$). Increasing field strength in the Taylor regime leads to stronger deformations in agreement with figure~\ref{fig:steadytaylor}. Interestingly, the transition to electrorotation breaks this trend and leads to a relaxation of the drop towards a more spherical shape. This decrease in $\mathcal{D}$ with the onset of rotation can be rationalized as a result of a change in the nature of the flow. In the Taylor regime, the axisymmetric toroidal vortex flow illustrated in figure~\ref{fig:snaptaylor} is dominated by straining and causes the elongation of the drop in the equatorial plane; under Quincke rotation, the flow becomes primarily rotational and therefore has a weaker effect on drop shape. This qualitative change also has an impact on the charge distribution, which is much smoother in the Quincke regime than in the Taylor regime, thus reducing the effective dipole and the magnitude of electric stresses at a given field magnitude.

\begin{figure}
\centering
\includegraphics[width=\linewidth]{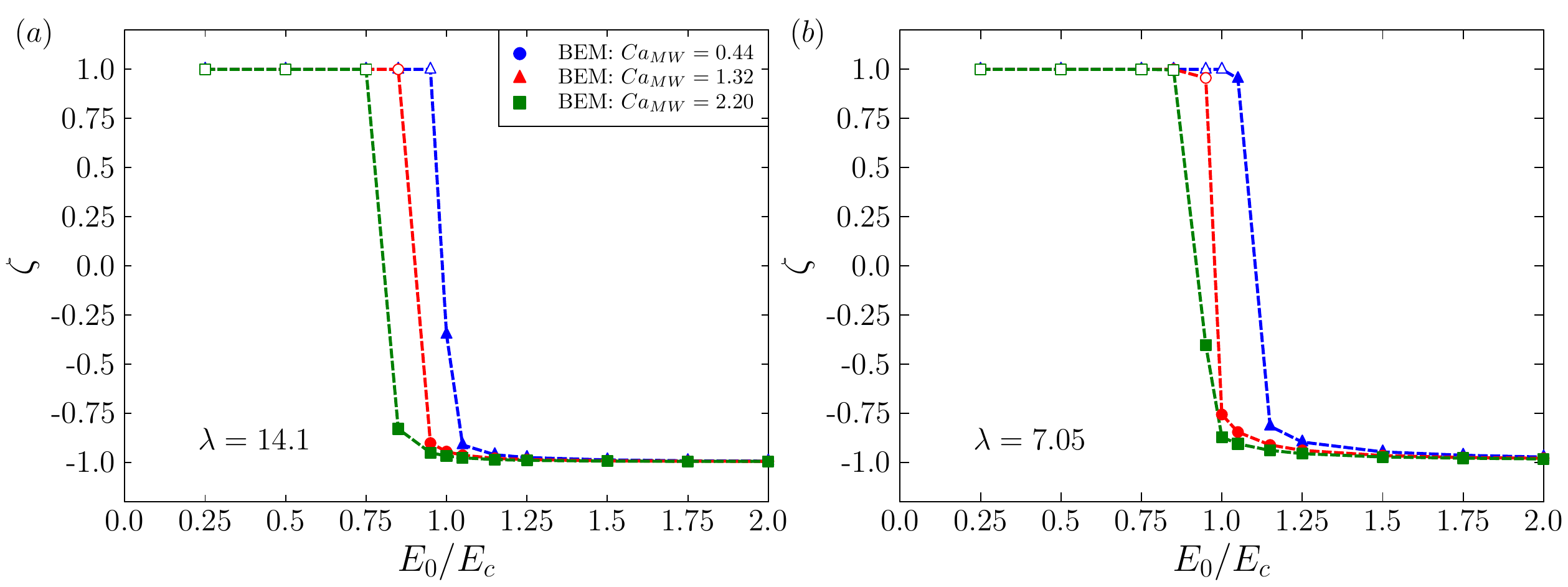}
\caption{(Color online) Parameter $\zeta$, defined in equation~(\ref{eq:zeta}) and calculated at the position of the drop centroid for system 2c,  as a function of electric field strength for (\textit{a}) $\lambda=14.1$ and (\textit{b}) $\lambda = 7.05$. Values of $\zeta$ close to $1$ or $-1$ describe flows dominated by either strain or rotation, respectively. }
\label{fig:zeta}
\end{figure}

In order to quantify more precisely the nature of the flow inside the drop, we introduce a parameter $\zeta$ as \vspace{-0.2cm}
\begin{align}
\zeta = \frac{\mathrm{tr}({\mathsfbi{S}^2}) - \mathrm{tr}({\mathsfbi{W}^2})}{\mathrm{tr}({\mathsfbi{S}^2}) + \mathrm{tr}({\mathsfbi{W}^2})}, \label{eq:zeta}
\end{align}
where $\mathsfbi{S}=\tfrac{1}{2}(\boldsymbol{\nabla}\boldsymbol{v} + \boldsymbol{\nabla}\boldsymbol{v}^T)$ and   $\mathsfbi{W}=\tfrac{1}{2}(\boldsymbol{\nabla}\boldsymbol{v} - \boldsymbol{\nabla}\boldsymbol{v}^T)$ denote the rate-of-strain and rate-of-rotation tensors, respectively, which we evaluate at the centroid of the drop. With this definition, values of $\zeta$ close to $+1$ and $-1$ describe flows dominated by strain and rotation, respectively. The dependence of $\zeta$ on electric field strength in the steady state is shown in figure~\ref{fig:zeta} for different values of $Ca_{MW}$ and for two viscosity ratios. In the Taylor regime, $\zeta=1$ at the center of the drop, which is to be expected for the axisymmetric Taylor flow. As the transition to electrorotation takes place, $\zeta$ rapidly jumps to a value close to $-1$, which indicates a drastic change in the nature of the flow. Note, however, that $\zeta$ is not strictly $-1$ in the Quincke regime, implying that the flow retains a straining component; nonetheless, we find that $\zeta\rightarrow -1$ as $E_0/E_c$ keeps increasing and the rotational component of the flow becomes more dominant.

\section{Concluding remarks}\label{sec:conclusion}

In this work, we have developed a three-dimensional boundary element method for the unsteady electrohydrodynamics of a deformable viscous drop based on the complete Melcher--Taylor leaky dielectric model including nonlinear charge convection. Our method extends previous numerical studies in this field \citep{sherwood1988,baygents1998,lac2007,lanauze2015}, which either were restricted to axisymmetric shapes or neglected charge convection. Our results were first shown to reproduce the steady oblate and prolate shapes known to arise in the Taylor regime of weak fields and compared favorably with previous models and experiments. In stronger fields, the experimentally observed symmetry-breaking bifurcation and transition to Quincke electrorotation was also captured for the first time in simulations. A phase diagram for the transition between the two regimes was constructed, and the evolution of drop shape and tilt angle with increasing field strength was discussed and shown to agree well with experiments. Our numerical simulations also allowed us to characterize the nature of the flow, which is not easily visualized experimentally, and demonstrated a transition from a strain-dominated flow in the Taylor regime to a primarily rotational flow in the Quincke regime. 

Our simulations, which were limited to isolated viscous drops in moderate electric fields, open the way for the study of more complex situations. The cases of very strong fields and low-viscosity drops remain challenging numerically: our numerical method was found to become unstable in these limits, thus preventing us from investigating the unsteady chaotic dynamics observed in the experiments of \cite{salipante2013}. Another difficulty arising in this case is the formation of charge shocks as shown by previous studies \citep{lanauze2013,das2016} and illustrated in figure~\ref{fig:snaptaylor}. The accurate treatment of these sharp charge discontinuities should require the implementation of a shock capturing scheme for the solution of the charge conservation equation. High-order weighted essentially non-oscillatory (WENO) schemes \citep{hu1999} within a finite-volume formulation could prove useful towards this purpose, though their implementation on unstructured meshes is non-trivial. 

Extensions of the present work could also include the consideration of sedimentation, which couples nonlinearly with the electrohydrodynamic problem as a result of charge convection and was recently discussed theoretically in the limit of small deformations and weak fields \citep{bandopadhyay2016,yarivalmog16}. Droplet-droplet and droplet-wall interactions, either pairwise or in collections of multiple drops, would also be interesting to analyze in the light of recent experiments on droplet pairs \citep{dommer16} and emulsions \citep{varshney2012,varshney2016}. Such interactions also have  yet to be studied numerically, which would likely requires the use of an accelerated algorithm such as the fast multipole method \citep{zinchenko2000}.


\section*{Acknowledgements}
The authors thank P.~Vlahovska and P.~Salipante for helpful discussions and suggestions, A.~Khair and J.~Lanauze for useful comments and sharing their experimental data, A.~Spann and M.~Theillard for discussions on the implementation of the numerical scheme. Acknowledgement is made to the Donors of the American Chemical Society Petroleum Research Fund for support of this research through grant 53240-ND9.

\appendix
\section{Discrete surface parametrization}\label{app:surfaceparam}

The drop surface is divided into $N_\triangle$ six-node curved triangular elements, allowing for computation of local curvature. Each physical three-dimensional element is mapped to an isosceles right triangle residing in a plane parametrized by coordinates $s_1$ and $s_2$. Any point  $\boldsymbol{x}$ inside the element in the physical space is represented by means of six basis functions $\phi_i$ that are defined on each triangle, exact expressions for which are provided by \citet{pozrikidis1992,pozrikidis2002}:\vspace{-0.2cm}
\begin{equation}
\boldsymbol{x} = \sum_{i=1}^6 \boldsymbol{x}_i \phi_i(s_1,s_2). \label{eq:xpar} \vspace{-0.1cm}
\end{equation}
Similarly, any scalar, vectorial or tensorial field $f(\boldsymbol{x})$ can be represented as 
\begin{equation}
f(\boldsymbol{x})=\sum_{i=1}^{6}f_{i} \phi_{i}(s_1,s_2). \label{eq:functionpar} 
\end{equation}
The unit tangent vectors in the directions of $s_1$ and $s_2$ in physical space are obtained as
\begin{equation}
\boldsymbol{e}_{s_1} =\sum_{i=1}^6 \boldsymbol{x}_i \frac{\partial \phi_i}{\partial s_1}, ~~~ \boldsymbol{e}_{s_2} =\sum_{i=1}^6 \boldsymbol{x}_i \frac{\partial \phi_i}{\partial s_2},
\end{equation}
from which the unit normal vector is found as
\begin{equation}
\boldsymbol{n} = \frac{1}{h_S} \boldsymbol{e}_{s_1} \times \boldsymbol{e}_{s_2},
\end{equation}
where $h_S(\xi,\eta) = |\boldsymbol{e}_{s_1} \times \boldsymbol{e}_{s_2}|$ is the surface metric.
We define the metric tensor $\mathsfbi{A}$ as
\begin{equation}
A_{ij} = \frac{\partial x_k}{\partial s_i} \frac{\partial x_k}{\partial s_j},
\end{equation}
which allows us to find the surface divergence of any surface vector $\boldsymbol{v}(\boldsymbol{x})$ as
\begin{equation}
\boldsymbol{\nabla}_s \boldsymbol{\cdot} \boldsymbol{v} = A^{-1}_{ij} \frac{\partial v_k}{\partial s_i} \frac{\partial x_k}{\partial s_j}. \label{eq:curvature}
\end{equation}
In particular, we use equation \eqref{eq:curvature} to compute both the total curvature $2\kappa_m =\boldsymbol{\nabla}_s \boldsymbol{\cdot} \boldsymbol{n}$ and charge convection term $\boldsymbol{\nabla}_s \boldsymbol{\cdot} (q\boldsymbol{v})$. Since these terms are computed locally in each triangular element, the value of these quantities at a global node that is shared between multiple elements is obtained by averaging the values at the local nodes. An alternative method of computing the surface divergence of a vector consists in using the Stokes theorem, which yields
\begin{equation}
\kappa_m = \frac{1}{2S_E}\int_{S_E} \boldsymbol{\nabla}_s \cdot \boldsymbol{v} \,\mathrm{d}S = \frac{1}{2S_E} \oint_{C_E} \boldsymbol{b} \times \boldsymbol{v} \,\mathrm{d}\ell,
\end{equation}
where $\boldsymbol{b}=\boldsymbol{t}\times\boldsymbol{n}$ is the outward unit normal to the edges of the triangular element and $S_E$ and $C_E$ are the element area and contour, respectively \citep{pozrikidis2011}. The Stokes theorem also forms the basis of the finite volume method for the charge conservation equation. We did not find any significant difference between these two methods and the curvature is computed using \eqref{eq:curvature} in this work.

Given the representation of equation (\ref{eq:functionpar}), surface integrals of field variables  are simply obtained by analytical quadrature of the basis functions. The surface gradient $\bnabla_{s}f(\boldsymbol{x})=(\mathsfbi{I}-\boldsymbol{nn})\bcdot\bnabla f(\boldsymbol{x})$ of a field variable can also be determined by solving a $3\times 3$ linear system at each quadrature point on the mesh:
\begin{equation}
\frac{\partial \boldsymbol{x}}{\partial s_1}\bcdot\bnabla_s f=\frac{\partial f}{\partial s_1}, \quad \frac{\partial \boldsymbol{x}}{\partial s_2}\bcdot\bnabla_s f=\frac{\partial f}{\partial s_2}, \quad \boldsymbol{n}\bcdot\bnabla_s f=0, 
\end{equation}
where the partial derivatives are calculated analytically by differentiation of equations (\ref{eq:xpar}) and (\ref{eq:functionpar}).

\section{Wielandt's deflation technique}\label{app:Wielandt}

We present Wielandt's deflation technique, which is employed for faster convergence of the iterative GMRES solver used for solving the Stokes boundary integral equation. The dimensionless integral equation for the interfacial velocity reads
\begin{align}
\begin{split}
\boldsymbol{v} (\boldsymbol{x}_0)  &+ \frac{\lambda-1}{8 \pi}\oint_S [\boldsymbol{v}(\boldsymbol{x}) - \boldsymbol{v}(\boldsymbol{x}_0)]\boldsymbol{\cdot} \mathsfbi{T}(\boldsymbol{x};\boldsymbol{x}_0) \boldsymbol{\cdot} \boldsymbol{n} (\boldsymbol{x}) \,\mathrm{d}S(\boldsymbol{x}) \\
& = - \frac{1}{8\pi Ma}\oint_S \llbracket \boldsymbol{f}^H(\boldsymbol{x})\rrbracket \boldsymbol{\cdot} \mathsfbi{G} (\boldsymbol{x};\boldsymbol{x}_0) \,\mathrm{d}S(\boldsymbol{x}). \label{eq:illcond}
\end{split}
\end{align}
Wielandt's deflation technique consists in formulating a different boundary integral equation in terms of an auxiliary field $\boldsymbol{w}$, which is obtained after removal of rigid body motion and uniform expansion solutions \citep{pozrikidis1992}:
\begin{align}
\begin{split}
& \boldsymbol{w} (\boldsymbol{x}_0) + \frac{(\lambda-1)}{8\pi} \left[ \oint_S [\boldsymbol{w}(\boldsymbol{x}) - \boldsymbol{w}(\boldsymbol{x}_0)] \boldsymbol{\cdot} \mathsfbi{T} (\boldsymbol{x};\boldsymbol{x}_0) \boldsymbol{\cdot} \boldsymbol{n}(\boldsymbol{x}) \,\mathrm{d}S(\boldsymbol{x})+4\pi \boldsymbol{w}^\prime(\boldsymbol{x}_0) \right. \\
& \left.   - \frac{4\pi}{S} \boldsymbol{n}(\boldsymbol{x}_0) \oint_S \boldsymbol{w}(\boldsymbol{x}) \boldsymbol{\cdot} \boldsymbol{n}(\boldsymbol{x}) \,\mathrm{d}S(\boldsymbol{x}) \right] = -\frac{1}{8\pi Ma} \oint_S \llbracket \boldsymbol{f}^H(\boldsymbol{x})\rrbracket  \boldsymbol{\cdot} \mathsfbi{G}(\boldsymbol{x};\boldsymbol{x}_0)\,\mathrm{d}S(\boldsymbol{x}), \label{eq:wielandt1}
\end{split}
\end{align}
where $\boldsymbol{w}^\prime$ denotes the rigid body motion:
\begin{align}
\boldsymbol{w}^\prime (\boldsymbol{x}_0) =  \boldsymbol{U} + \boldsymbol{\Omega} \times (\boldsymbol{x}_0 -\boldsymbol{x}_c). 
\end{align}
Here, $\boldsymbol{x}_{c}$ is the surface centroid, and $\boldsymbol{U}$ and  $\boldsymbol{\Omega}$ are the translational and rotational velocities, respectively:
\begin{align}
&\boldsymbol{x}_c = \frac{1}{S} \oint_S \boldsymbol{x} \,\mathrm{d}S(\boldsymbol{x}), \\
&\boldsymbol{U}= \frac{1}{S} \oint_S \boldsymbol{w}(\boldsymbol{x}) \,\mathrm{d}S(\boldsymbol{x}), \\
&\boldsymbol{\Omega} = \mathsfbi{M}^{-1} \cdot \oint_S (\boldsymbol{x} -\boldsymbol{x}_c) \times \boldsymbol{w}(\boldsymbol{x}) \,\mathrm{d}S(\boldsymbol{x}),
\end{align}
where the matrix $\mathsfbi{M}$ is given by
\begin{equation}
\mathsfbi{M} = \oint_S \left[\mathsfbi{I}|\boldsymbol{x}-\boldsymbol{x}_c|^2 - (\boldsymbol{x}-\boldsymbol{x}_c)(\boldsymbol{x}-\boldsymbol{x}_c)\right] \,\mathrm{d}S(\boldsymbol{x}).
\end{equation}
Substituting these expressions into (\ref{eq:wielandt1}) yields the desired integral equation for $\boldsymbol{w}$, which no longer suffers from the ill-conditioning of (\ref{eq:illcond}):
\begin{align}
\begin{split}
&\boldsymbol{w}(\boldsymbol{x}_0) + \frac{(\lambda-1)}{8\pi} \left[ \oint_S [\boldsymbol{w}(\boldsymbol{x}) - \boldsymbol{w}(\boldsymbol{x}_0)]\boldsymbol{\cdot} \mathsfbi{T}(\boldsymbol{x};\boldsymbol{x}_0) \boldsymbol{\cdot} \boldsymbol{n}(\boldsymbol{x}) \,\mathrm{d}S(\boldsymbol{x}) \right. \\
+&\left. \frac{4\pi}{S} \oint_S \boldsymbol{w}(\boldsymbol{x}) \,\mathrm{d}S(\boldsymbol{x}) + 4\pi\left(\mathsfbi{M}^{-1} \cdot \oint_S (\boldsymbol{x} -\boldsymbol{x}_c) \times \boldsymbol{w}(\boldsymbol{x}) \,\mathrm{d}S(\boldsymbol{x}) \right) \times (\boldsymbol{x}_0 -\boldsymbol{x}_c) \right. \\
-&\left.\frac{4\pi}{S} \boldsymbol{n}(\boldsymbol{x}_0) \oint_S \boldsymbol{w}(\boldsymbol{x}) \boldsymbol{\cdot} \boldsymbol{n}(\boldsymbol{x}) \,\mathrm{d}S(\boldsymbol{x}) \right] = -\frac{1}{8\pi Ma} \oint_S \llbracket \boldsymbol{f}^H(\boldsymbol{x})\rrbracket \boldsymbol{\cdot} \mathsfbi{G}(\boldsymbol{x};\boldsymbol{x}_0) \,\mathrm{d}S(\boldsymbol{x}).
\end{split}
\end{align}
Having determined the auxiliary field $\boldsymbol{w}$, we compute the actual interfacial velocity as
\begin{align}
\boldsymbol{v} = \boldsymbol{w} + \frac{\lambda-1}{2} \boldsymbol{w}^\prime.
\end{align}

\bibliographystyle{jfm}
\bibliography{bemquincke}

\begin{thebibliography}{79}
\expandafter\ifx\csname natexlab\endcsname\relax\def\natexlab#1{#1}\fi
\def\au#1{#1} \def\ed#1{#1} \def\yr#1{#1}\def\at#1{#1}\def\jt#1{\textit{#1}}
  \def\bt#1{#1}\def\bvol#1{\textbf{#1}} \def\vol#1{#1} \def\pg#1{#1}
  \def\publ#1{#1}\def\arxiv#1{#1}\def\org#1{#1}\def\st#1{\textit{#1}}

\bibitem[Ajayi(1978)]{ajayi1978}
{\sc \au{Ajayi, O.~O.}} \yr{1978}  \at{A note on {T}aylor's electrohydrodynamic
  theory}.  \jt{Proc. R. Soc. Lond. A}  \bvol{364},  \pg{499--507}.

\bibitem[Allan \& Mason(1962)]{allan1962}
{\sc \au{Allan, R.~S.} \& \au{Mason, S.~G.}} \yr{1962}  \at{Particle behaviour
  in shear and electric fields. {I}. {D}eformation and burst of fluid drops}.
  \jt{Proc. R. Soc. Lond. A}  \bvol{267},  \pg{45--61}.

\bibitem[Bandopadhyay {\em et~al.\/}(2016)Bandopadhyay, Mandal, Kishore \&
  Chakraborty]{bandopadhyay2016}
{\sc \au{Bandopadhyay, A.}, \au{Mandal, S.}, \au{Kishore, N.~K.} \&
  \au{Chakraborty, S.}} \yr{2016}  \at{Uniform electric-field-induced lateral
  migration of a sedimenting drop}.  \jt{J. Fluid Mech.}  \bvol{792},
  \pg{553--589}.

\bibitem[Basaran {\em et~al.\/}(2013)Basaran, Gao \& Bhat]{basaran2013}
{\sc \au{Basaran, O.~A.}, \au{Gao, H.} \& \au{Bhat, P.~P.}} \yr{2013}
  \at{Nonstandard inkjets}.  \jt{Annu. Rev. Fluid Mech.}  \bvol{45},
  \pg{85--113}.

\bibitem[Baygents {\em et~al.\/}(1998)Baygents, Rivette \& Stone]{baygents1998}
{\sc \au{Baygents, J.~C.}, \au{Rivette, N.~J.} \& \au{Stone, H.~A.}} \yr{1998}
  \at{Electrohydrodynamic deformation and interaction of drop pairs}.  \jt{J.
  Fluid Mech.}  \bvol{368},  \pg{359--375}.

\bibitem[Bjorklund(2009)]{bjorklund2009}
{\sc \au{Bjorklund, E.}} \yr{2009}  \at{The level-set method applied to droplet
  dynamics in the presence of an electric field}.  \jt{Comput. Fluids}
  \bvol{38},  \pg{358--369}.

\bibitem[Blanchard(1963)]{blanchard1963}
{\sc \au{Blanchard, D.~C.}} \yr{1963}  \at{The electrification of the
  atmosphere by particles from bubbles in the sea}.  \jt{Prog. Oceanogr.}
  \bvol{1},  \pg{73--202}.

\bibitem[Brazier-Smith(1971)]{brazier1971a}
{\sc \au{Brazier-Smith, P.~R.}} \yr{1971}  \at{Stability and shape of isolated
  and pairs of water drops in an electric field}.  \jt{Phys. Fluids}
  \bvol{14},  \pg{1--6}.

\bibitem[Brazier-Smith {\em et~al.\/}(1971)Brazier-Smith, Jennings \&
  Latham]{brazier1971b}
{\sc \au{Brazier-Smith, P.~R.}, \au{Jennings, S.~G.} \& \au{Latham, J}}
  \yr{1971}  \at{An investigation of the behaviour of drops and drop-pairs
  subjected to strong electrical forces}.  \jt{Proc. R. Soc. Lond. A}
  \bvol{325},  \pg{363--376}.

\bibitem[Castellanos(2014)]{castellanos2014}
{\sc \au{Castellanos, A.}} \yr{2014} {\em Electrohydrodynamics\/}.
  \publ{Springer}.

\bibitem[Das \& Saintillan(2013)]{das2013}
{\sc \au{Das, D.} \& \au{Saintillan, D.}} \yr{2013}  \at{Electrohydrodynamic
  interaction of spherical particles under {Q}uincke rotation}.  \jt{Phys. Rev.
  E}  \bvol{87},  \pg{043014}.

\bibitem[Das \& Saintillan(2017)]{das2016}
{\sc \au{Das, D.} \& \au{Saintillan, D.}} \yr{2017}  \at{A nonlinear
  small-deformation theory for transient droplet electrohydrodynamics}.  \jt{J.
  Fluid Mech.}  \bvol{810},  \pg{225--253}.

\bibitem[Dommersnes {\em et~al.\/}(2016)Dommersnes, Mikkelsen \&
  Fossum]{dommer16}
{\sc \au{Dommersnes, P.}, \au{Mikkelsen, A.} \& \au{Fossum, J.}} \yr{2016}
  \at{Electro-hydrodynamic propulsion of counter-rotating pickering drops}.
  \jt{J. Eur. Phys. J. Spec. Top.}  \bvol{225},  \pg{699--706}.

\bibitem[Dubash \& Mestel(2007{\natexlab{{\em a\/}}})]{dubash2007a}
{\sc \au{Dubash, N.} \& \au{Mestel, A.~J.}} \yr{2007{\natexlab{{\em a\/}}}}
  \at{Behaviour of a conducting drop in a highly viscous fluid subject to an
  electric field}.  \jt{J. Fluid Mech}  \bvol{581},  \pg{469--493}.

\bibitem[Dubash \& Mestel(2007{\natexlab{{\em b\/}}})]{dubash2007b}
{\sc \au{Dubash, N.} \& \au{Mestel, A.~J.}} \yr{2007{\natexlab{{\em b\/}}}}
  \at{Breakup behavior of a conducting drop suspended in a viscous fluid
  subject to an electric field}.  \jt{Phys. Fluids}  \bvol{19},  \pg{072101}.

\bibitem[Eow \& Ghadiri(2002)]{eow2002}
{\sc \au{Eow, J.~S.} \& \au{Ghadiri, M.}} \yr{2002}  \at{Electrostatic
  enhancement of coalescence of water droplets in oil: a review of the
  technology}.  \jt{Chem. Eng. J.}  \bvol{85},  \pg{357--368}.

\bibitem[Esmaeeli \& Sharifi(2011)]{esmaeeli2011}
{\sc \au{Esmaeeli, A.} \& \au{Sharifi, P.}} \yr{2011}  \at{Transient
  electrohydrodynamics of a liquid drop}.  \jt{Phys. Rev. E}  \bvol{84},
  \pg{036308}.

\bibitem[Feng(1999)]{feng1999}
{\sc \au{Feng, J.~Q.}} \yr{1999}  \at{Electrohydrodynamic behaviour of a drop
  subjected to a steady uniform electric field at finite electric {R}eynolds
  number}.  \jt{Proc. R. Soc. Lond. A}  \bvol{455},  \pg{2245--2269}.

\bibitem[Feng(2002)]{feng2002}
{\sc \au{Feng, J.~Q.}} \yr{2002}  \at{A 2{D} electrohydrodynamic model for
  electrorotation of fluid drops}.  \jt{J. Colloid Interface Sci.}  \bvol{246},
   \pg{112--121}.

\bibitem[Feng \& Scott(1996)]{feng1996}
{\sc \au{Feng, J.~Q.} \& \au{Scott, T.~C.}} \yr{1996}  \at{A computational
  analysis of electrohydrodynamics of a leaky dielectric drop in an electric
  field}.  \jt{J. Fluid Mech.}  \bvol{311},  \pg{289--326}.

\bibitem[Ha \& Yang(2000{\natexlab{{\em a\/}}})]{ha2000a}
{\sc \au{Ha, J.-W.} \& \au{Yang, S.-M.}} \yr{2000{\natexlab{{\em a\/}}}}
  \at{Deformation and breakup of {N}ewtonian and non-{N}ewtonian conducting
  drops in an electric field}.  \jt{J. Fluid Mech.}  \bvol{405},
  \pg{131--156}.

\bibitem[Ha \& Yang(2000{\natexlab{{\em b\/}}})]{ha2000b}
{\sc \au{Ha, J.-W.} \& \au{Yang, S.-M.}} \yr{2000{\natexlab{{\em b\/}}}}
  \at{Electrohydrodynamics and electrorotation of a drop with fluid less
  conductive than that of the ambient fluid}.  \jt{Phys. Fluids}  \bvol{12},
  \pg{764--772}.

\bibitem[Harris \& O'Konski(1957)]{harris1957}
{\sc \au{Harris, F.~E.} \& \au{O'Konski, C.~T.}} \yr{1957}  \at{Dielectric
  properties of aqueous ionic solutions at microwave frequencies}.  \jt{J.
  Phys. Chem.}  \bvol{61},  \pg{310--319}.

\bibitem[Haywood {\em et~al.\/}(1991)Haywood, Renksizbulut \&
  Raithby]{haywood1991}
{\sc \au{Haywood, R.~J.}, \au{Renksizbulut, M.} \& \au{Raithby, G.~D.}}
  \yr{1991}  \at{Transient deformation of freely-suspended liquid droplets in
  electrostatic fields}.  \jt{AIChE J.}  \bvol{37},  \pg{1305--1317}.

\bibitem[He {\em et~al.\/}(2013)He, Salipante \& Vlahovska]{he2013}
{\sc \au{He, H.}, \au{Salipante, P.~F.} \& \au{Vlahovska, P.~M.}} \yr{2013}
  \at{Electrorotation of a viscous droplet in a uniform direct current electric
  field}.  \jt{Phys. Fluids}  \bvol{25},  \pg{032106}.

\bibitem[Hirata {\em et~al.\/}(2000)Hirata, Kikuchi, Tsukada \&
  Hozawa]{hirata2000}
{\sc \au{Hirata, T.}, \au{Kikuchi, T.}, \au{Tsukada, T.} \& \au{Hozawa, M.}}
  \yr{2000}  \at{Finite element analysis of electrohydrodynamic time-dependent
  deformation of dielectric drop under uniform {DC} electric field.}  \jt{J.
  Chem. Engng Japan}  \bvol{33},  \pg{160--167}.

\bibitem[Hu \& Shu(1999)]{hu1999}
{\sc \au{Hu, C.} \& \au{Shu, C.-W.}} \yr{1999}  \at{Weighted essentially
  non-oscillatory schemes on triangular meshes}.  \jt{J. Comput. Phys.}
  \bvol{150},  \pg{97--127}.

\bibitem[Hu {\em et~al.\/}(2015)Hu, Lai \& Young]{hu2015}
{\sc \au{Hu, W.-F.}, \au{Lai, M.-C.} \& \au{Young, Y.-N.}} \yr{2015}  \at{A
  hybrid immersed boundary and immersed interface method for
  electrohydrodynamic simulations}.  \jt{J. Comput. Phys.}  \bvol{282},
  \pg{47--61}.

\bibitem[Huang {\em et~al.\/}(2003)Huang, Zhang, Kotaki \&
  Ramakrishna]{huang2003}
{\sc \au{Huang, Z.-M.}, \au{Zhang, Y.-Z.}, \au{Kotaki, M.} \& \au{Ramakrishna,
  S.}} \yr{2003}  \at{A review on polymer nanofibers by electrospinning and
  their applications in nanocomposites}.  \jt{Compos. Sci. Technol.}
  \bvol{63},  \pg{2223--2253}.

\bibitem[Jaswon(1963)]{jaswon1963}
{\sc \au{Jaswon, M.~A.}} \yr{1963}  \at{Integral equation methods in potential
  theory. {I}}.  \jt{Proc. R. Soc. Lond. A}  \bvol{275},  \pg{23--32}.

\bibitem[Jones(1984)]{jones1984}
{\sc \au{Jones, T.B.}} \yr{1984}  \at{Quincke rotation of spheres}.  \jt{IEEE
  Transactions on Industry Applications}  \bvol{IA-20},  \pg{845--849}.

\bibitem[Karyappa {\em et~al.\/}(2014)Karyappa, Deshmukh \&
  Thaokar]{karyappa2014}
{\sc \au{Karyappa, R.~B.}, \au{Deshmukh, S.~D.} \& \au{Thaokar, R.~M.}}
  \yr{2014}  \at{Breakup of a conducting drop in a uniform electric field}.
  \jt{J. Fluid Mech.}  \bvol{754},  \pg{550--589}.

\bibitem[Kennedy {\em et~al.\/}(1994)Kennedy, Pozrikidis \&
  Skalak]{kennedy1994}
{\sc \au{Kennedy, M.~R.}, \au{Pozrikidis, C.} \& \au{Skalak, R.}} \yr{1994}
  \at{Motion and deformation of liquid drops, and the rheology of dilute
  emulsions in simple shear flow}.  \jt{Computer Fluids}  \bvol{23},
  \pg{251--278}.

\bibitem[Kim \& Karrila(2013)]{kim2013}
{\sc \au{Kim, S.} \& \au{Karrila, S.~J.}} \yr{2013} {\em Microhydrodynamics:
  Principles and Selected Applications\/}.  \publ{Dover}.

\bibitem[Krause \& Chandratreya(1998)]{krause1998}
{\sc \au{Krause, S.} \& \au{Chandratreya, P.}} \yr{1998}  \at{Electrorotation
  of deformable fluid droplets}.  \jt{J. Colloid Interface Sci.}  \bvol{206},
  \pg{10--18}.

\bibitem[Lac \& Homsy(2007)]{lac2007}
{\sc \au{Lac, E.} \& \au{Homsy, G.~M.}} \yr{2007}  \at{Axisymmetric deformation
  and stability of a viscous drop in a steady electric field}.  \jt{J. Fluid
  Mech.}  \bvol{590},  \pg{239--264}.

\bibitem[Lanauze {\em et~al.\/}(2013)Lanauze, Walker \& Khair]{lanauze2013}
{\sc \au{Lanauze, J.~A.}, \au{Walker, L.~M.} \& \au{Khair, A.~S.}} \yr{2013}
  \at{The influence of inertia and charge relaxation on electrohydrodynamic
  drop deformation}.  \jt{Phys. Fluids}  \bvol{25},  \pg{112101}.

\bibitem[Lanauze {\em et~al.\/}(2015)Lanauze, Walker \& Khair]{lanauze2015}
{\sc \au{Lanauze, J.~A.}, \au{Walker, L.~M.} \& \au{Khair, A.~S.}} \yr{2015}
  \at{Nonlinear electrohydrodynamics of slightly deformed oblate drops}.
  \jt{J. Fluid Mech.}  \bvol{774},  \pg{245--266}.

\bibitem[Landau {\em et~al.\/}(1984)Landau, Lifshitz \&
  Pitaevski{\`i}]{landau1984}
{\sc \au{Landau, L.~D.}, \au{Lifshitz, E.~M.} \& \au{Pitaevski{\`i}, L.~P.}}
  \yr{1984} {\em Electrodynamics of continuous media\/}.  \publ{Elsevier}.

\bibitem[Laser \& Santiago(2004)]{laser2004}
{\sc \au{Laser, D.~J.} \& \au{Santiago, J.~G.}} \yr{2004}  \at{A review of
  micropumps}.  \jt{J. Micromech. Microengng.}  \bvol{14},  \pg{R35}.

\bibitem[Loewenberg \& Hinch(1996)]{loewenberg1996}
{\sc \au{Loewenberg, M.} \& \au{Hinch, E.~J.}} \yr{1996}  \at{Numerical
  simulation of a concentrated emulsion in shear flow}.  \jt{J. Fluid Mech.}
  \bvol{321},  \pg{395--419}.

\bibitem[L{\'o}pez-Herrera {\em et~al.\/}(2011)L{\'o}pez-Herrera, Popinet \&
  Herrada]{lopez2011}
{\sc \au{L{\'o}pez-Herrera, J.~M.}, \au{Popinet, S.} \& \au{Herrada, M.~A.}}
  \yr{2011}  \at{A charge-conservative approach for simulating
  electrohydrodynamic two-phase flows using volume-of-fluid}.  \jt{J. Comput.
  Phys.}  \bvol{230},  \pg{1939--1955}.

\bibitem[Melcher \& Taylor(1969)]{melcher1969}
{\sc \au{Melcher, J.~R.} \& \au{Taylor, G.~I.}} \yr{1969}
  \at{Electrohydrodynamics: a review of the role of interfacial shear
  stresses}.  \jt{Annu. Rev. Fluid Mech.}  \bvol{1},  \pg{111--146}.

\bibitem[Miksis(1981)]{miksis1981}
{\sc \au{Miksis, M.~J.}} \yr{1981}  \at{Shape of a drop in an electric field}.
  \jt{Phys. Fluids}  \bvol{24},  \pg{1967--1972}.

\bibitem[O'Konski \& Thacher(1953)]{konski1953}
{\sc \au{O'Konski, C.~T.} \& \au{Thacher, H.~C.}} \yr{1953}  \at{The distortion
  of aerosol droplets by an electric field}.  \jt{J. Phys. Chem.}  \bvol{57},
  \pg{955--958}.

\bibitem[Pozrikidis(1992)]{pozrikidis1992}
{\sc \au{Pozrikidis, C.}} \yr{1992} {\em Boundary Integral and Singularity
  Methods for Linearized Viscous Flow\/}.  \publ{Cambridge University Press}.

\bibitem[Pozrikidis(2002)]{pozrikidis2002}
{\sc \au{Pozrikidis, C.}} \yr{2002} {\em A Practical Guide to Boundary Element
  Methods with the Software Library BEMLIB\/}.  \publ{CRC Press}.

\bibitem[Pozrikidis(2011)]{pozrikidis2011}
{\sc \au{Pozrikidis, C.}} \yr{2011} {\em Introduction to Theoretical and
  Computational Fluid Dynamics\/}.  \publ{Oxford University Press}.

\bibitem[Quincke(1896)]{quincke1896}
{\sc \au{Quincke, G.}} \yr{1896}  \at{{\"Uber} rotationen im constanten
  electrischen {Felde}}.  \jt{Ann. Phys. Chem.}  \bvol{59},  \pg{417}.

\bibitem[Rallison \& Acrivos(1978)]{rallison1978}
{\sc \au{Rallison, J.~M.} \& \au{Acrivos, A.}} \yr{1978}  \at{A numerical study
  of the deformation and burst of a viscous drop in an extensional flow}.
  \jt{J. Fluid Mech.}  \bvol{89},  \pg{191--200}.

\bibitem[Saad \& Schultz(1986)]{saad1986}
{\sc \au{Saad, Y.} \& \au{Schultz, M.~H.}} \yr{1986}  \at{{GMRES}: A
  generalized minimal residual algorithm for solving nonsymmetric linear
  systems}.  \jt{SIAM J. Sci. Stat. Comput.}  \bvol{7},  \pg{856--869}.

\bibitem[Salipante \& Vlahovska(2010)]{salipante2010}
{\sc \au{Salipante, P.~F.} \& \au{Vlahovska, P.~M.}} \yr{2010}
  \at{Electrohydrodynamics of drops in strong uniform dc electric fields}.
  \jt{Phys. Fluids}  \bvol{22},  \pg{112110}.

\bibitem[Salipante \& Vlahovska(2013)]{salipante2013}
{\sc \au{Salipante, P.~F.} \& \au{Vlahovska, P.~M.}} \yr{2013}
  \at{Electrohydrodynamic rotations of a viscous droplet}.  \jt{Phys. Rev. E}
  \bvol{88},  \pg{043003}.

\bibitem[Sato {\em et~al.\/}(2006)Sato, Kaji, Mochizuki \& Mori]{sato2006}
{\sc \au{Sato, H.}, \au{Kaji, N.}, \au{Mochizuki, T.} \& \au{Mori, Y.~H.}}
  \yr{2006}  \at{Behavior of oblately deformed droplets in an immiscible
  dielectric liquid under a steady and uniform electric field}.  \jt{Phys.
  Fluids}  \bvol{18},  \pg{127101}.

\bibitem[Saville(1997)]{saville1997}
{\sc \au{Saville, D.~A.}} \yr{1997}  \at{Electrohydrodynamics: the
  {T}aylor--{M}elcher leaky dielectric model}.  \jt{Annu. Rev. Fluid Mech.}
  \bvol{29},  \pg{27--64}.

\bibitem[Schnitzer \& Yariv(2015)]{schnitzer2015}
{\sc \au{Schnitzer, O.} \& \au{Yariv, E.}} \yr{2015}  \at{The
  {T}aylor-{M}elcher leaky dielectric model as a macroscale electrokinetic
  description}.  \jt{J. Fluid Mech.}  \bvol{773},  \pg{1--33}.

\bibitem[Schramm(1992)]{schramm1992}
{\sc \au{Schramm, L.~L.}} \yr{1992} {\em Emulsions: Fundamentals and
  Applications in the Petroleum Industry\/}.  \publ{American Chemical Society}.

\bibitem[Sellier(2006)]{sellier2006}
{\sc \au{Sellier, A.}} \yr{2006}  \at{On the computation of the derivatives of
  potentials on a boundary by using boundary-integral equations}.  \jt{Comput.
  Methods Appl. Mech. Eng.}  \bvol{196},  \pg{489--501}.

\bibitem[Sherwood(1988)]{sherwood1988}
{\sc \au{Sherwood, J.~D.}} \yr{1988}  \at{Breakup of fluid droplets in electric
  and magnetic fields}.  \jt{J. Fluid Mech.}  \bvol{188},  \pg{133--146}.

\bibitem[Shkadov \& Shutov(2002)]{shkadov02}
{\sc \au{Shkadov, V.~Y.} \& \au{Shutov, A.~A.}} \yr{2002}  \at{Drop and bubble
  deformation in an electric field}.  \jt{Fluid Dyn.}  \bvol{37},
  \pg{713--724}.

\bibitem[Simpson(1909)]{simpson1909}
{\sc \au{Simpson, G.~C.}} \yr{1909}  \at{On the electricity of rain and its
  origin in thunderstorms}.  \jt{Phil. Trans. R. Soc. Lond. A}  \bvol{209},
  \pg{379--413}.

\bibitem[Stone {\em et~al.\/}(2004)Stone, Stroock \& Ajdari]{stone2004}
{\sc \au{Stone, H.~A.}, \au{Stroock, A.~D.} \& \au{Ajdari, A.}} \yr{2004}
  \at{Engineering flows in small devices: microfluidics toward a
  lab-on-a-chip}.  \jt{Annu. Rev. Fluid Mech.}  \bvol{36},  \pg{381--411}.

\bibitem[Supeene {\em et~al.\/}(2008)Supeene, Koch \&
  Bhattacharjee]{supeene2008}
{\sc \au{Supeene, G.}, \au{Koch, C.~R.} \& \au{Bhattacharjee, S.}} \yr{2008}
  \at{Deformation of a droplet in an electric field: {N}onlinear transient
  response in perfect and leaky dielectric media}.  \jt{J. Colloid Interface
  Sci.}  \bvol{318},  \pg{463--476}.

\bibitem[Symm(1963)]{symm1963}
{\sc \au{Symm, G.~T.}} \yr{1963}  \at{Integral equation methods in potential
  theory. {II}}.  \jt{Proc. R. Soc. Lond. A}  \bvol{275},  \pg{33--46}.

\bibitem[Taylor(1964)]{taylor1964}
{\sc \au{Taylor, G.~I.}} \yr{1964}  \at{Disintegration of water drops in an
  electric field}.  \jt{Proc. R. Soc. Lond. A}  \bvol{280},  \pg{383--397}.

\bibitem[Taylor(1966)]{taylor1966}
{\sc \au{Taylor, G.~I.}} \yr{1966}  \at{Studies in electrohydrodynamics. {I}.
  {T}he circulation produced in a drop by electrical field}.  \jt{Proc. R. Soc.
  Lond. A}  \bvol{291},  \pg{159--166}.

\bibitem[Taylor(1969)]{taylor1969}
{\sc \au{Taylor, G.~I.}} \yr{1969}  \at{Electrically driven jets}.  \jt{Proc.
  R. Soc. Lond. A}  \bvol{313},  \pg{453--475}.

\bibitem[Torza {\em et~al.\/}(1971)Torza, Cox \& Mason]{torza1971}
{\sc \au{Torza, S.}, \au{Cox, R.~G.} \& \au{Mason, S.~G.}} \yr{1971}
  \at{Electrohydrodynamic deformation and burst of liquid drops}.  \jt{Phil.
  Trans. R. Soc. Lond. A}  \bvol{269},  \pg{295--319}.

\bibitem[Tsukada {\em et~al.\/}(1993)Tsukada, Katayama, Ito \& M.]{tsukada93}
{\sc \au{Tsukada, T.}, \au{Katayama, T.}, \au{Ito, Y.} \& \au{M., Hozawa}}
  \yr{1993}  \at{Theoretical and experimental studies of circulations inside
  and outside a deformed drop under a uniform electric field}.  \jt{J. Chem.
  Engng Japan}  \bvol{26},  \pg{698--703}.

\bibitem[Varshney {\em et~al.\/}(2012)Varshney, Ghosh, Bhattacharya \&
  Yethiraj]{varshney2012}
{\sc \au{Varshney, A.}, \au{Ghosh, S.}, \au{Bhattacharya, S.} \& \au{Yethiraj,
  A.}} \yr{2012}  \at{Self organization of exotic oil-in-oil phases driven by
  tunable electrohydrodynamics}.  \jt{Sci. Rep.}  \bvol{2},  \pg{738}.

\bibitem[Varshney {\em et~al.\/}(2016)Varshney, Gohil, Sathe, RV, Joshi,
  Bhattacharya, Yethiraj \& Ghosh]{varshney2016}
{\sc \au{Varshney, A.}, \au{Gohil, S.}, \au{Sathe, M.}, \au{RV, S.~R.},
  \au{Joshi, J.~B.}, \au{Bhattacharya, S.}, \au{Yethiraj, A.} \& \au{Ghosh,
  S.}} \yr{2016}  \at{Multiscale flow in an electro-hydrodynamically driven
  oil-in-oil emulsion}.  \jt{Soft Matter}  \bvol{12},  \pg{1759--1764}.

\bibitem[Vizika \& Saville(1992)]{vizika1992}
{\sc \au{Vizika, O} \& \au{Saville, D.~A.}} \yr{1992}  \at{The
  electrohydrodynamic deformation of drops suspended in liquids in steady and
  oscillatory electric fields}.  \jt{J. Fluid Mech.}  \bvol{239},  \pg{1--21}.

\bibitem[Vlahovska(2016)]{vlahovska2016}
{\sc \au{Vlahovska, P.~M.}} \yr{2016}  \at{Electrohydrodynamic instabilities of
  viscous drops}.  \jt{Phys. Rev. Fluids}  \bvol{1},  \pg{060504}.

\bibitem[Yariv \& Almog(2016)]{yarivalmog16}
{\sc \au{Yariv, E.} \& \au{Almog, Y.}} \yr{2016}  \at{The effect of
  surface-charge convection on the settling velocity of spherical drops in a
  uniform electric field}.  \jt{J. Fluid Mech.}  \bvol{797},  \pg{536--548}.

\bibitem[Yariv \& Frankel(2016)]{yariv2016}
{\sc \au{Yariv, E.} \& \au{Frankel, I.}} \yr{2016}  \at{Electrohydrodynamic
  rotation of drops at large electric reynolds numbers}.  \jt{J. Fluid Mech.}
  \bvol{788},  \pg{R2}.

\bibitem[Yon \& Pozrikidis(1998)]{yon1998}
{\sc \au{Yon, S.} \& \au{Pozrikidis, C.}} \yr{1998}  \at{A
  finite-volume/boundary-element method for flow past interfaces in the
  presence of surfactants, with application to shear flow past a viscous drop}.
   \jt{Computers Fluids}  \bvol{27},  \pg{879--902}.

\bibitem[Zhang {\em et~al.\/}(2013)Zhang, Zahn \& Lin]{zhang2013}
{\sc \au{Zhang, J.}, \au{Zahn, J.~D.} \& \au{Lin, H.}} \yr{2013}  \at{Transient
  solution for droplet deformation under electric fields}.  \jt{Phys. Rev. E}
  \bvol{87},  \pg{043008}.

\bibitem[Zinchenko \& Davis(2000)]{zinchenko2000}
{\sc \au{Zinchenko, A.~Z.} \& \au{Davis, R.~H.}} \yr{2000}  \at{An efficient
  algorithm for hydrodynamical interaction of many deformable drops}.  \jt{J.
  Comput. Phys.}  \bvol{157},  \pg{539--587}.

\bibitem[Zinchenko {\em et~al.\/}(1997)Zinchenko, Rother \&
  Davis]{zinchenko1997}
{\sc \au{Zinchenko, A.~Z.}, \au{Rother, M.~A.} \& \au{Davis, R.~H.}} \yr{1997}
  \at{A novel boundary-integral algorithm for viscous interaction of deformable
  drops}.  \jt{Phys. Fluids}  \bvol{9},  \pg{1493--1511}.

\end{thebibliography}

\end{document}